\documentclass[aps,twocolumn,superscriptaddress,notitlepage,amsmath,amssymb]{revtex4-2}

\usepackage{hyperref}
\usepackage{mathtools}
\usepackage{esdiff}
\usepackage{xcolor}
\usepackage[per-mode=symbol]{siunitx}

\newcommand{\e}[1]{\mathrm{e}^{#1}}
\newcommand{\im}{\mathrm{i}}
\newcommand{\dif}{\mathrm{d}}

\DeclarePairedDelimiter{\paren}{\lparen}{\rparen}
\DeclarePairedDelimiter{\sbrak}{\lbrack}{\rbrack}
\DeclarePairedDelimiter{\cbrak}{\lbrace}{\rbrace}
\DeclarePairedDelimiter{\abs}{\lvert}{\rvert}

\DeclarePairedDelimiter{\mean}{\langle}{\rangle}
\DeclarePairedDelimiterX{\var}[1]{\langle}{\rangle}{\delta#1^2}
\DeclarePairedDelimiterX{\covar}[2]{\langle}{\rangle}{\delta#1\delta#2}

\DeclareMathOperator{\RE}{Re}

\DeclareMathOperator{\diag}{diag}
\DeclareMathOperator{\sinc}{sinc}
\DeclareMathOperator{\sech}{sech}

\begin{document}
\title{Quantum noise dynamics in nonlinear pulse propagation}

\author{Edwin~Ng}
\thanks{These authors contributed equally to this work.}
\affiliation{Physics \& Informatics Laboratories, NTT Research, Inc., Sunnyvale, California 94085, USA}
\affiliation{E.\,L.\,Ginzton Laboratory, Stanford University, Stanford, California 94305, USA}

\author{Ryotatsu~Yanagimoto}
\thanks{These authors contributed equally to this work.}
\affiliation{Physics \& Informatics Laboratories, NTT Research, Inc., Sunnyvale, California 94085, USA}
\affiliation{E.\,L.\,Ginzton Laboratory, Stanford University, Stanford, California 94305, USA}
\affiliation{School of Applied and Engineering Physics, Cornell University, Ithaca, New York 14853, USA}

\author{Marc~Jankowski}
\affiliation{Physics \& Informatics Laboratories, NTT Research, Inc., Sunnyvale, California 94085, USA}
\affiliation{E.\,L.\,Ginzton Laboratory, Stanford University, Stanford, California 94305, USA}

\author{M.\,M.\,Fejer}
\affiliation{E.\,L.\,Ginzton Laboratory, Stanford University, Stanford, California 94305, USA}

\author{Hideo~Mabuchi}
\affiliation{E.\,L.\,Ginzton Laboratory, Stanford University, Stanford, California 94305, USA}

\date{\today}

\begin{abstract}
The propagation of ultrafast pulses in dispersion-engineered waveguides, exhibiting strong field confinement in both space and time, is a promising avenue towards single-photon nonlinearities in an all-optical platform.
However, quantum engineering in such systems requires new numerical tools and physical insights to harness their complicated multimode and nonlinear quantum dynamics.
In this work, we use a self-consistent, multimode Gaussian-state model to capture the nonlinear dynamics of broadband quantum fluctuations and correlations, including entanglement.
Notably, despite its parametrization by Gaussian states, our model exhibits nonlinear dynamics in both the mean field and the quantum correlations, giving it a marked advantage over conventional linearized treatments of quantum noise, especially for systems exhibiting gain saturation and strong nonlinearities.
Numerically, our approach takes the form of a Gaussian split-step Fourier (GSSF) method, naturally generalizing highly efficient SSF methods used in classical ultrafast nonlinear optics; the equations for GSSF evaluate in $\mathcal O(M^2\log M)$ time for an $M$-mode system with $\mathcal O(M^2)$ quantum correlations.
To demonstrate the broad applicability of GSSF, we numerically study quantum noise dynamics and multimode entanglement in several ultrafast systems, from canonical soliton propagation in third-order ($\chi^{(3)}$) waveguides to saturated $\chi^{(2)}$ broadband parametric generation and supercontinuum generation, e.g., as recently demonstrated in thin-film lithium niobate nanophotonics.
\end{abstract}

\maketitle

\section{Introduction}

The concept of shot noise is a prevailing paradigm for understanding fundamental quantum fluctuations in electromagnetic radiation, arising from the discrete nature of photons~\cite{Clerk2010}.
As photonic devices push towards the ultimate limits of energy efficiency, however, quantum fluctuations become an increasingly ubiquitous and limiting factor in their operation, and the potential emergence of non-Poissonian and correlated photon statistics---e.g., squeezing~\cite{Walls1983}, photon (anti-)bunching~\cite{Hong1987}, and quantum diffusion of optical pulses~\cite{Gordon1986,Bao2021}---necessitates a more sophisticated treatment of quantum noise.
At the same time, properly harnessing such nonclassical phenomena presents major opportunities in photonics research, with applications from quantum-enhanced metrology~\cite{Giovannetti2011,Pezze2008,LIGO2013,Ozeki2020} to quantum information processing~\cite{Zhong2020,Asavarant2019,Zhang2014,Arrazola2021,Takeda2019}.

An emerging but promising approach for accessing this nonclassical regime is the use of dispersion-engineered nonlinear nanophotonics~\cite{Jankowski2021-review}, where the spatial~\cite{Lu2020,Zhao2022} and temporal confinement of light to ultrashort pulses propagating in sub-wavelength waveguides significantly enhances the nonlinear polarization produced per photon.
In principle, such devices can access single-photon nonlinearities~\cite{Yanagimoto2022-temporal}, in which full quantum models are needed to describe photon correlations~\cite{Yanagimoto2021_mps,Gilchrist1997,Drummond2014}.
However, even as experimental efforts advance towards this critical milestone, many transitional and practically important devices, from high-gain parametric amplifiers~\cite{Nehra2022,Kashiwazaki2020,Vahlbruch2007} to low-power microcombs~\cite{Liu2018}, are expected to operate in a more intermediate, semiclassical regime, where it suffices to account for first- and second-order (i.e., Gaussian) correlations in the quantum fluctuations.
A complete understanding of these leading-order quantum effects is also vital for navigating the classical-quantum transition, allowing us to conceptually interpolate between these regimes and facilitating the development of hybrid semiclassical-quantum models~\cite{Yanagimoto2021-non-gaussian,Yanagimoto2021-spie}.
The systematic treatment of Gaussian quantum correlations has recently been formalized into the language of Gaussian-state quantum optics~\cite{Weedbrook2012,Olivares2012,Braunstein2005}, a framework that describes the action of basic linear components such as squeezers, beamspitters, phaseshifters, etc., as discrete Gaussian operations on Gaussian states.

From a classical perspective, however, the dynamics of light are much richer than a Gaussian-state formalism based on discrete operations might suggest.
Broadband fields evolving under nonlinear partial differential equations prescribed by Maxwell's equations, e.g., the Lugiato-Lefever equation~\cite{Lugiato1987} and the nonlinear Schr\"odinger equation~\cite{Zakharov1972}, can support a rich phenomenology of emergent multimode dynamics, such as rogue waves~\cite{Meng2021,Tlidi2022}, chaos, and solitons~\cite{Kivshar2003,Kivshar1998}, enabling breakthrough technologies such as optical frequency (micro-)combs~\cite{Kippenberg2018,Grelu2012,Herr2014} in the process.
As quantum fluctuations become increasingly relevant to the operation of highly multimode and nonlinear devices, we require a unified framework~\cite{Yanagimoto2021-spie} that leverages the mathematical efficacy of multimode Gaussian-state models while capturing the physical expressivity of nonlinear ultrafast dynamics.
The demand is especially acute for guiding the development of emerging platforms like thin-film lithium niobate (TFLN) nanophotonics~\cite{Zhang2017,Jankowski2021-review,Nehra2022}, which is already anticipating a regime of attojoule-level, femtosecond nonlinear optics in next-generation devices.

In this paper, we show how the framework of multimode Gaussian quantum optics~\cite{Quesada2022} can be integrated with the nonlinear dynamics of ultrafast pulse propagation, allowing us to study the roles quantum fluctuations play even in current-generation devices.
Our approach is a natural Gaussian-state generalization of the classical split-step Fourier (SSF) method used in nonlinear ultrafast optics, modified to systematically treat multimode quantum noise, correlations and entanglement on the same dynamical footing as the mean field, without the use of ad hoc noise models or Monte-Carlo techniques.
In contrast to conventional linearized treatments such as undepleted-pump approximations~\cite{Hosaka2016,Haus1990,Helt2020}, our Gaussian SSF (GSSF) approach uses a self-consistent Gaussian-state approximation to the quantum dynamics~\cite{Schack1990,Verstraelen2018,Verstraelen2020,Huang2022,Navarrete-Benlloch2014} to take into account nonlinear corrections to the mean-field dynamics induced by quantum fluctuations.
These corrections are necessary to ensure energy conservation in the high-efficiency, low-energy regimes of nonlinear nanophotonics, where saturation energies are orders of magnitude lower than in bulk or fiber optics and linearized models are often inadequate.

We apply our method to numerically study the dynamics of quantum noise in several illustrative examples from nonlinear ultrafast optics.
Using the GSSF version of the nonlinear Schr\"odinger equation, we study the canonical Kerr soliton and show how multimode quantum fluctuations can destabilize the classical waveform.
We also look at optical parametric generation~\cite{Jankowski2022, Ledezma2022}, in which intense squeezing of a signal pulse results in pump depletion solely due to parametric fluorescence~\cite{Florez2020, Xing2022, Kinsler1993}.
Finally, we simulate supercontinuum generation based on broadband, saturated second-harmonic generation~\cite{Jankowski2020, Jankowski2021} and analyze how quantum entanglement of the octave-spanning frequency comb affects the quantum noise limit for the detection of carrier-envelope-offset beat notes in $f-2f$ interferometry.
Notably, the latter two examples involve device parameters that have already been demonstrated experimentally using TFLN waveguides~\cite{Jankowski2021-review}, underscoring the utility of our GSSF framework for engineering ultrafast quantum nonlinear devices.

\section{Gaussian approximation of nonlinear dynamics} \label{sec:single-mode}

To illustrate our scheme and compare it with other approaches, we consider one of the simplest nonlinear optical models, the \emph{single-mode} Kerr Hamiltonian $\hat H_\text{Kerr} \coloneqq \frac12 \hbar g \hat a^{\dagger2} \hat a^2$ (note we use the $\coloneqq$ notation for definitions, and $\partial_y x \coloneqq \dif x / \dif y$).
Physically, $\hat H_\text{Kerr}$ can be seen as describing a single trapped mode in a cavity experiencing self-phase modulation, and the Heisenberg equation of motion for its quantum dynamics is $\im\partial_{gt} \hat{a}=\hat{a}^\dagger\hat{a}^2$.
Note that such a single-mode model does not inherit the modeling challenges intrinsic to multimode quantum dynamics of pulse propagation, and thus, it should be seen as only a toy model in the context of this work. Nevertheless, as we show in this section, we can still obtain useful insights translatable to generic multimode scenarios through the studies of the single-mode toy model.

Our approach is to assume that the system can be well described by a Gaussian state characterized by the mean $\mean{\hat a}$ and covariances $\covar{\hat a^\dagger}{\hat a}$ and $\var{\hat a}$, where $\delta\hat a \coloneqq \hat a - \mean{\hat a}$ is the fluctuation operator corresponding to $\hat a$.
Intuitively, the mean corresponds to (or generalizes) the classical field, while the covariances describe the statistics of the system's \emph{quantum noise}.
Of course, since $\hat H_\text{Kerr}$ is a nonlinear Hamiltonian, the dynamics in principle can generate non-Gaussian features in the state.
Here, we are primarily interested in a systematic approach for neglecting such non-Gaussian features in order to arrive at a \emph{Gaussian approximation} of the dynamics, which physically is well justified outside regimes of strong single-photon nonlinearities.

To derive the equations of motion for the mean field, we take expectations on both sides of the Heisenberg equation of motion, to obtain
\begin{align} \label{eq:1mode-quantum-mean}
\im\partial_{gt} \mean{\hat a} = \mean{\hat a^\dagger \hat a^2}.
\end{align}
The righthand side involves an expectation over a higher-order product of operators, for which we require a suitable approximation.
We now describe one conventional approach for dealing with this problem, which we call the \emph{linearized treatment}.
We then generalize this treatment with a nonlinear Gaussian model to include nonlinear corrections.

\begin{figure}[ht]
\centering
\includegraphics[width=0.5\textwidth]{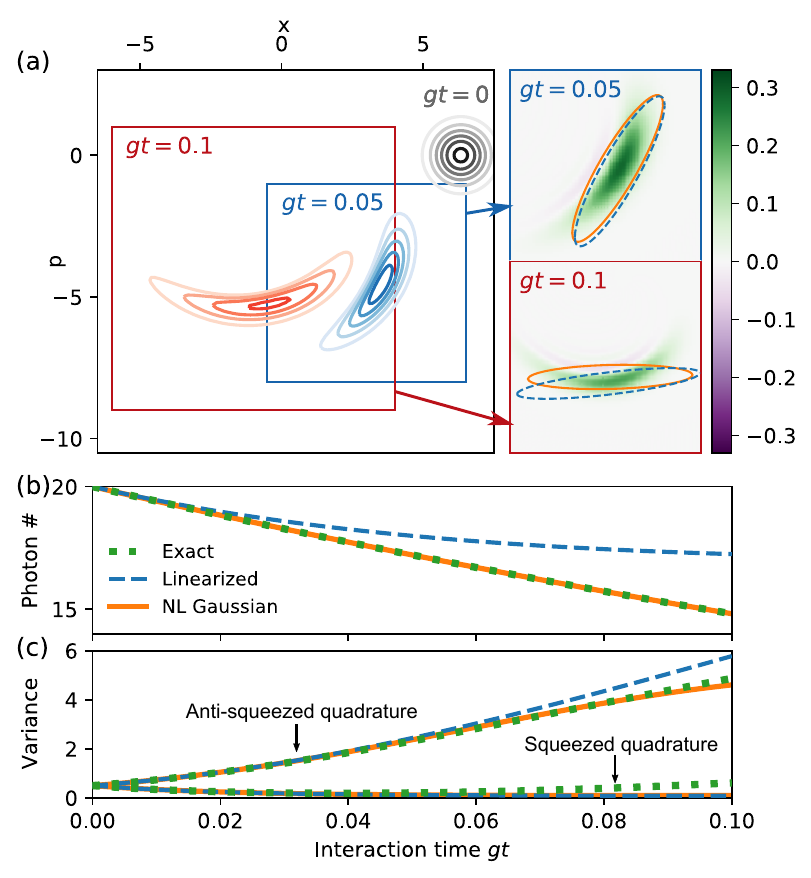}
\caption{
Illustrative example in approximating quantum fluctuations in a single-mode Kerr model by Gaussian quantum noise.
(a) Phase-space portraits of a coherent state evolving under the single-mode Kerr Hamiltonian $\hat{H}_\text{Kerr}$, with zoomed-in Wigner functions shown inset at various times.
We compare two different approaches for approximating the Gaussian moments of the fluctuations (see main text); the linearized treatment is shown with a dash-dotted ellipse, while the nonlinear (NL) Gaussian model is shown solid.
As a function of time, we also compare between the two models total photon number (b) and major- and minor-axis variances of the quantum fluctuations (c).
For reference, the initial coherent state has displacement $\mean{\hat a} = \sqrt{\num{20}}$ and we assume a linear field decay rate of $\kappa = 1.5g$ (see Appendix~\ref{sec:dissipation} for treating loss in the NL Gaussian model).
Full-quantum simulations are performed using QuantumOptics.jl~\cite{Kraemer2018}.
} \label{fig:fig1}
\end{figure}

\subsection{Linearized treatment}

In the linearized treatment, we make two main assumptions to simplify \eqref{eq:1mode-quantum-mean}.
First, we assume the state is well approximated by a coherent state, so we can write
\begin{align} \label{eq:coherent-factorization}
\mean{\hat a^\dagger \hat a^2} \mapsto \mean{\hat a^\dagger} \mean{\hat a}^2.
\end{align}
As a result, we immediately recover the (classical) mean-field equation of motion for a Kerr cavity,
\begin{align} \label{eq:1mode-linear-mean}
\im\partial_{gt} \mean{\hat a} = \mean{\hat a^\dagger} \mean{\hat a}^2,
\end{align}
which can now be solved without employing any knowledge of how the quantum noise evolves in the system.

Second, for the covariance equations, we discard any terms on the righthand side which are second-or-higher-order in the fluctuation operators.
This generates a \emph{linearized} equation of motion for the fluctuation operator
\begin{align} \label{eq:1mode-linear-fluctuation}
\im\partial_{gt}\delta\hat{a} = \mean{\hat a}^2 \delta\hat a^\dagger + 2\mean{\hat a^\dagger} \mean{\hat a} \delta\hat a.
\end{align}
From this, it follows that the covariances also evolve linearly, according to
\begin{subequations} \label{eq:1mode-linear-covars}
\begin{align}
\im\partial_{gt}\var{\hat a} &= \mean{\hat a}^2 \mean{\delta\hat a^\dagger \delta\hat a + \delta\hat a\delta\hat a^\dagger} + 4\mean{\hat a^\dagger} \mean{\hat a} \var{\hat a} \\
\im\partial_{gt}\covar{\hat a^\dagger}{\hat a} &= \mean{\hat a}^2 \mean{\delta\hat{a}^{\dagger2}} - \mean{\hat a^\dagger}^2 \var{\hat a}, \label{eq:1mode-linear-covars-2}
\end{align}
\end{subequations}
which, together with \eqref{eq:1mode-linear-mean}, constitute the dynamics under the linearized treatment of $\hat H_\text{Kerr}$.
In fact, we can analytically solve these equations: For an initial coherent state with $\mean{\hat a} = \alpha_0$, we have for the mean
$\mean{\hat a} = \alpha_0 \e{-\im\tau}$ and for the covariances
\begin{align}
\var{\hat a} = -\e{-2\im\tau}(\tau^2+\im\tau),
\qquad
\covar{\hat a^\dagger}{\hat a} = \tau^2,
\end{align}
where we have defined $\tau \coloneqq \abs{\alpha_0}^2 gt$.

In the linearized treatment, there is an asymmetry or separation of scales between the classical and semiclassical dynamics: While the evolution of the quantum fluctuations are driven by the evolution of the mean field, the mean itself evolves purely classically and is unaffected by the quantum noise.
This inherent inconsistency is often acceptable in situations where a very large mean field is required to produce even modest amounts of squeezing, but it can lead to unphysical consequences, such as violation of photon-number (energy) conservation, in more mesoscopic regimes of operation.
In this case, the mean photon number under the linearized dynamics is
\begin{equation}
\bar n \coloneqq \abs{\mean{\hat a}}^2 + \covar{\hat a^\dagger}{\hat a} = \abs{\alpha_0}^2(1 + \tau),
\end{equation}
which is clearly increasing with time. 

In Fig.~\ref{fig:fig1}, we show the evolution of an initial coherent state under $\hat{H}_\text{Kerr}$ comparing the exact quantum dynamics (dotted line) to the linearized treatment (dashed lines).
We see that linearization overestimates both the photon number and the variances of the quantum fluctuations, especially at later times.
As hinted in the figure, however, such issues can be mitigated by turning to a self-consistent nonlinear Gaussian model.

\subsection{Nonlinear Gaussian model}

Using again the simple single-mode $\hat H_\text{Kerr}$, we now outline the essential ingredients for an alternative approach based on a self-consistent Gaussian-state approximation.
Our goal is again to derive equations of motion for the mean $\mean{\hat a}$ and the covariances $\covar{\hat a^\dagger}{\hat a}$ and $\var{\hat a}$, but, here, we keep all terms in intermediate calculations and only apply Gaussian-state assumption at the end, after expanding higher-order moments as needed.

Rather than \eqref{eq:coherent-factorization}, we instead have the exact expression
\begin{equation}
\mean{\hat a^\dagger \hat a^2} = \mean{\hat a^\dagger}\mean{\hat a}^2 + 2\mean{\hat a}\covar{\hat a^\dagger}{\hat a} + \mean{\hat a^\dagger}\var{\hat a} + \mean{\delta a^\dagger \delta\hat a^2}.
\end{equation}
With this, the mean-field equation of motion becomes
\begin{align} \label{eq:1mode-gaussian-mean}
\im\partial_{gt}\mean{\hat a} = \mean{\hat a^\dagger} \mean{\hat a}^2 + 2\mean{\hat a}\langle\delta\hat a^\dagger\delta\hat a\rangle + \mean{\hat a^\dagger}\var{\hat a},
\end{align}
where the only approximation we have made is that $\mean{\delta\hat a^\dagger \delta\hat a^2} = 0$, which is necessarily true for a Gaussian state since this term is an odd-order central moment.
In contrast to \eqref{eq:1mode-linear-mean}, the equation of motion for the mean now involves the covariances.

To obtain the equation of motion for the covariances, we also first obtain the dynamics of the fluctuations.
However, without making the linearization approximation, the exact form of \eqref{eq:1mode-linear-fluctuation} is instead
\begin{align}
\im\partial_{gt} \delta\hat a &= \mean{\hat a}^2 \delta\hat a^\dagger + 2\mean{\hat a^\dagger}\mean{\hat a}\delta\hat a + \delta\hat a^\dagger \delta\hat a^2 \label{eq:1mode-gaussian-fluctuation} \\
&\qquad{} + 2\mean{\hat a}\paren{\delta\hat a^\dagger\delta\hat a - \covar{\hat a^\dagger}{\hat a}} + \mean{\hat a^\dagger}\paren{\delta\hat a^2 - \var{\hat a}}. \nonumber
\end{align}
We next utilize the chain rule, taking care to preserve operator ordering, via
\begin{equation}
\partial_t\covar{\hat z_1}{\hat z_1} = \mean{(\partial_t\delta\hat z_1)(\delta\hat z_2)} + \mean{\delta\hat z_1 (\partial_t\delta\hat z_2)}.
\end{equation}
Applying this to, e.g., the equation of motion for $\var{\hat a}$,
\begin{align*}
\im\partial_{gt}\var{\hat a} &= \mean{\hat a}^2 \covar{\hat a^\dagger}{\hat a} + \mean{\hat a}^2 \covar{\hat a}{\hat a^\dagger} + 4\mean{\hat a^\dagger}\mean{\hat a}\var{\hat a} \nonumber\\
&\qquad{}+ \mean{\delta\hat a^\dagger \delta\hat a^3} + \mean{\delta\hat a \delta\hat a^\dagger \delta\hat a^2},
\end{align*}
where the only approximation we have made is again the elimination of odd central moments, i.e., the terms stemming from the second line of \eqref{eq:1mode-gaussian-fluctuation}.
Compared to \eqref{eq:1mode-linear-covars}, we see that there are fourth-order correction terms~\footnote{
Interestingly, it turns out in that in the single-mode case, such higher-order corrections only affect the equation for $\var{\hat a}$, and in fact \eqref{eq:1mode-linear-covars-2} is unchanged.
However, in the general multimode case, higher-order moments enter into the evolution of all non-diagonal moments in general.
}.

Finally, we require one additional step in order to simplify the fourth-order moments occurring in the second line above, as this term is not generally zero for a Gaussian state.
However, for a Gaussian state, it turns out that even higher-order central moments can be \emph{decomposed} into sums of products of covariances only.
In particular, for this case, we can use the expansion~\footnote{
More generally, for a Gaussian state, $\mean{\delta\hat z_1 \cdots \delta\hat z_n} = \sum_{p \in \mathbb P_n} \prod_{(i,j) \in p} \covar{\hat z_i}{\hat z_j}$, where $\mathbb P_n$ denotes the set of all order-preserving pair partitions of $\{1, \ldots, n\}$.
For example, for $n = 4$, the elements of $\mathbb P_4$ are the 3 pair partitions $\{(1,2),(3,4)\}$, $\{(1,3),(2,4)\}$, and $\{(1,4),(2,3)\}$.
Taking these elements in the sum-of-products produces \eqref{eq:fourth-order-expansion}.
}
\begin{align} \label{eq:fourth-order-expansion}
&\mean{\delta\hat z_1 \delta\hat z_2 \delta\hat z_3 \delta\hat z_4}
= \covar{\hat z_1}{\hat z_2}\covar{\hat z_3}{\hat z_4} \\
&\qquad{}+ \covar{\hat z_1}{\hat z_3}\covar{\hat z_2}{\hat z_4}
+ \covar{\hat z_1}{\hat z_4}\covar{\hat z_2}{\hat z_3}. \nonumber
\end{align}
With this decomposition, we can now show that
\begin{subequations} \label{eq:1mode-gaussian-covars}
\begin{align}
\im\partial_{gt} \var{\hat a} &= \mean{\hat a^2} \mean{\delta\hat a^\dagger \delta\hat a + \delta\hat a \delta\hat a^\dagger} + 4\mean{\hat a^\dagger\hat a}\var{\hat a} \\
\im\partial_{gt}\covar{\hat a^\dagger}{\hat a} &= \mean{\hat a^2} \mean{\delta\hat a^{\dagger2}} - \mean{\hat a^{\dagger2}} \var{\hat a},
\end{align}
\end{subequations}
which, together with \eqref{eq:1mode-gaussian-mean}, constitute the nonlinear Gaussian model for $\hat H_\text{Kerr}$.
Note that in \eqref{eq:1mode-gaussian-covars}, we use the shorthands $\mean{\hat a^2} = \mean{\hat a}^2 + \var{\hat a}$ and $\mean{\hat a^\dagger \hat a} = \mean{\hat a^\dagger} \mean{\hat a} + \covar{\hat a^\dagger}{\hat a}$ to highlight similarities to the linearized treatment \eqref{eq:1mode-linear-covars}.
At the same time, we clearly see the distinction as well: The linearized treatment effectively assumes that $\var{\hat a} \ll \mean{\hat a}^2$ and $\covar{\hat a^\dagger}{\hat a} \ll |\mean{\hat a}|^2$ in evaluating the evolution of the variances.

We emphasize that the coupled equations \eqref{eq:1mode-gaussian-mean} and \eqref{eq:1mode-gaussian-covars} are \emph{nonlinear} differential equations, describing nonlinear evolution of the Gaussian moments.
As first pointed out in Ref.~\cite{Schack1990}, such models, while approximate, can capture a wider set of physical behaviors than linearized approximations, where the Gaussian moments follow strictly linear dynamics.
To distinguish the two, we therefore refer to such models as \emph{nonlinear Gaussian-state models}.
We also note that, in the single-mode case, these nonlinear equations are consistent with those derived using a similar approach in Ref.~\cite{Verstraelen2018}.

Finally, we can show that, in contrast to the linearized treatment, the nonlinear Gaussian-state model preserves photon number.
This can be seen by computing
\begin{equation}
\partial_t \bar n = \mean{\hat a^\dagger} (\partial_t\mean{\hat a}) + (\partial_t\mean{\hat a^\dagger}) \mean{\hat a} + \partial_t \covar{\hat a^\dagger}{\hat a} = 0.
\end{equation}
In fact, as shown in Fig.~\ref{fig:fig1}, the nonlinear Gaussian-state model exactly tracks the photon number of the correct quantum model, while providing a more faithful estimate of the variances compared to the linearized treatment.

\section{Quantum noise propagation in a chi(3) waveguide} \label{sec:chi3-formalism}

Extending this method to the broadband and multimode setting, we now consider a 1D waveguide with a non-dispersive third-order $\chi^{(3)}$ nonlinearity.
Theoretically, a continuum treatment of such a system can be quantized by introducing field operators $\hat\psi_z$ which obey continuum commutation relations $\sbrak[\big]{\hat\psi_z,\hat\psi_{z'}^\dagger} = \delta(z-z')$ and annihilate the quantized photon-polariton field of the medium at some spatial position $z$.
Using these field operators together with their Fourier duals $\hat\Psi_k \coloneqq \int \dif z\, \e{-\im kz} \hat\psi_z$, we consider a $\chi^{(3)}$ Hamiltonian
\begin{align}
\label{eq:chi3-hamiltonian-continuum}
\hat H_{\chi^{(3)}} &\coloneqq \hat H_\text{4wm} + \hat H_\text{lin}, \quad\text{where} \\
\hat H_\text{4wm} &\coloneqq \frac{\hbar}{2} \int \dif z\, g\, \hat\psi^{\dagger2}_z \hat\psi_z^2,
\quad
\hat H_\text{lin} \coloneqq \hbar \int \frac{\dif k}{2\pi}\, \Omega(k) \hat\Psi_k^\dagger \hat\Psi_k. \nonumber
\end{align}
Here, $g$ is a coupling rate related to the nonlinearity, while $\Omega(k)$ describes the linear dispersion of the field.
In this work, we are interested in the copropagating envelope of a pulse but not the carrier nor the absolute group velocity.
Thus, if $\omega(k_0+k)$ is the bare frequency of a monochromatic mode $\hat\Psi_k$ with wavevector offset by $k$ from the carrier's at $k_0$, then we define $\Omega(k) \coloneqq \omega(k_0+k) - \sbrak[\big]{\omega(k_0) + k\omega'(k_0)}$.
That is, we interpret $\hat\Psi_k$ in a frame rotating at $\omega(k_0+k) - \Omega(k) = \omega(k_0) + k\omega'(k_0)$, and $\hat\psi_z$ acts on a relative position $z$ comoving at $\omega'(k_0)$.

The Heisenberg equation of motion for $\hat\psi_z$ generated by $\hat H_{\chi^{(3)}}$ is
\begin{equation} \label{eq:nlse-quantum}
\im\partial_t \hat\psi_z = g \hat\psi_z^\dagger \hat\psi_z^2 + \Omega(-\im\partial_z)\hat\psi_z.
\end{equation}
The mean-field version of this equation, obtained by formally replacing $\hat\psi_z$ with a c-number function $\psi(z)$, is is the usual classical equation of motion for a mean-field waveform $\psi(z)$ in a $\chi^{(3)}$ waveguide with linear dispersion, of which the famous nonlinear Schr\"odinger equation (NLSE) is a special case when $\Omega$ is expanded to second order~\cite{Agrawal2019}.

The two terms representing the $\chi^{(3)}$ nonlinearity and linear dispersion above are respectively generated by the two Hamiltonians $\hat H_\text{4wm}$ and $\hat H_\text{lin}$.
They each take a simple local form in \eqref{eq:chi3-hamiltonian-continuum}, only when respectively expressed in position and momentum space, which are Fourier dual to one another; consequently, $\hat H_\text{4wm}$ and $\hat H_\text{lin}$ do not commute in general.
In numerical methods, it is well-known that such a situation can be effectively treated with a split-operator approach: Instead of trying to evolve the system under both Hamiltonians simultaneously, we Trotterize the dynamics by iteratively applying the evolution due to $\hat H_\text{3wm}$ and $\hat H_\text{lin}$ separately, using the Fourier transform to convert between position and momentum space as needed.
To facilitate this approach, we rewrite \eqref{eq:nlse-quantum} in terms of differential (super)operators
\begin{subequations} \label{eq:chi3-split-eoms}
\begin{align}
\dot{\mathcal N} \hat\psi_z &\coloneqq \frac{\im}{\hbar g} \sbrak{\hat H_\text{3wm}, \hat\psi_z} = -\im \hat\psi_z^\dagger\hat\psi_z^2, \label{eq:chi3-nonlinear-eom} \\
\dif\mathcal D \hat\Psi_k &\coloneqq \frac{\im}{\hbar} \sbrak{\hat H_\text{lin}, \hat\Psi_k} \,\dif t = -\im\Omega(k) \hat\Psi_k \,\dif t, \label{eq:chi3-linear-eom}
\end{align}
\end{subequations}
where we use the dot notation $\dot{\mathcal N} = \dif\mathcal N/\dif(gt)$ to denote differential evolution with respect to normalized time $gt$ in the nonlinear part, but we retain Leibniz notation $\dif\mathcal D = \dot{\mathcal D} \, \dif t$ in the linear evolution for convenience when treating loss as described in Appendix~\ref{sec:dissipation}.

Despite their superficial similarity to the classical model, both \eqref{eq:nlse-quantum} and \eqref{eq:chi3-split-eoms} are numerically intractable to solve directly.
Physically, the problem amounts to solving for the dynamics of an entire \emph{quantum field}, where each field degree of freedom (i.e., mode) occupies a bosonic Fock space.
Even if we discretize the field to $M$ modes and truncate the Fock space of each mode to $D$ dimensions (i.e., allowing at most $D-1$ photons per mode), the quantum state of the field lives in a $D^M$-dimensional (Hilbert) space, upon which operators such as $\hat H_\text{3wm}$ and $\hat H_\text{lin}$ act.
A typical discretization of the classical field $\psi(z)$ in the NSLE might employ $M = \num{1024}$ points, but even just allowing one photon per mode at $D = 2$, we have, at least without the use of sophisticated model reduction techniques, a $2^{1024}$-dimensional problem!

The situation becomes greatly simplified, however, if we are able to focus our attention solely on the Gaussian moments of the state, namely the mean $\mean{\hat\psi_z}$ (corresponding to an $M$-dimensional vector when discretized) and the covariances $\covar{\hat\psi_z}{\hat\psi_{z'}}$ and $\covar{\hat\psi^\dagger_z}{\hat\psi_{z'}}$ (each corresponding to an $M \times M$ matrix).
Thus, in a Gaussian framework, the numerical problem of solving for the quantum noise dynamics becomes $\mathcal O(M^2)$-dimensional.
As a result, our nonlinear Gaussian-state model has access to the same highly efficient numerical techniques employed by classical pulse propagation techniques, including the use of split-step methods based on the fast Fourier transform (FFT) and massively parallel computation on graphics processing units (GPUs).
Specifically, just as the cost of evolving the field over one time step for the classical NLSE is well known to be limited by FFT to $\mathcal O(M\log M)$, our method does the same for the full Gaussian moments of the field with only cost $\mathcal O(M^2\log M)$.
This makes our method a natural generalization of the classical split-step Fourier (SSF) method, and we therefore refer to our numerical approach, when applied to the problem of ultrafast pulse proagation, as a \emph{nonlinear Gaussian-state SSF (GSSF) method}.
In this work, we perform all GSSF simulations using a GPU implementation of the RK4IP split-step method~\cite{Hult2007} via the high-level Julia package CUDA.jl~\cite{Besard2018}.

As in Sec.~\ref{sec:single-mode}, the key contribution of this work is to prescribe nonlinear equations of motion for the mean and covariance of the multimode field $\hat\psi_z$, making only the assumption that the state is Gaussian.
Due to the split-step nature of the GSSF method, we have, as in the classical SSF, the additional requirement of applying the dispersive step due to $\hat H_\text{lin}$, but because \eqref{eq:chi3-linear-eom} is linear, we straightforwardly have
\begin{equation} \label{eq:chi3-gaussian-mean-linear}
\im\,\dif\mathcal D\mean{\hat\Psi_k} = \Omega(k) \mean{\hat\Psi_k} \,\dif t
\end{equation}
for the mean, and, for the covariances,
\begin{subequations} \label{eq:chi3-gaussian-covars-linear}
\begin{align}
\im\,\dif\mathcal D\covar{\hat\Psi_k}{\hat\Psi_{k'}} &= \paren[\big]{\Omega(k') + \Omega(k)} \covar{\hat\Psi_k}{\hat\Psi_{k'}} \,\dif t \\
\im\,\dif\mathcal D\covar{\hat\Psi^\dagger_k}{\hat\Psi_{k'}} &= \paren[\big]{\Omega(k') - \Omega(k)} \covar{\hat\Psi^\dagger_k}{\hat\Psi_{k'}} \,\dif t,
\end{align}
\end{subequations}
which can be analytically integrated.
Thus, the nontrivial part is deriving the equations of motion in the nonlinear (or real-space) step, but because \eqref{eq:chi3-nonlinear-eom} is local in $z$ (i.e., the differential evolution of $\hat\psi_z$ is decoupled from that of $\hat\psi_{z'}$ for $z \neq z'$), we can simply make use of the same methods already presented in Sec.~\ref{sec:single-mode} for the single-mode case, making sure to carefully track the multimode indices in the covariances.
For the mean, we have a modified version of the classical nonlinear step,
\begin{align} \label{eq:chi3-gaussian-mean-nonlin}
\im\,\dot{\mathcal N}\mean{\hat\psi_z} = \mean{\hat\psi^\dagger_z} \mean{\hat\psi_z}^2 + 2\mean{\hat\psi_z} \covar{\hat\psi^\dagger_z}{\hat\psi_z} + \mean{\hat\psi^\dagger_z} \var{\hat\psi_z},
\end{align}
where the last two terms are corrections due to coupling to the covariances.
Then the equations of motion for the covariances are, after some algebra,
\begin{subequations} \label{eq:chi3-gaussian-covars-nonlin}
\begin{align}
\im\,\dot{\mathcal N} &\covar{\hat\psi_z}{\hat\psi_{z'}}
= \mean{\hat\psi_z^2} \covar{\hat\psi_z^\dagger}{\hat\psi_{z'}}
+ \mean{\hat\psi_{z'}^2} \covar{\hat\psi_z}{\hat\psi^\dagger_{z'}} \nonumber\\
&\qquad{}+ 2 \paren[\big]{\mean{\hat\psi_z^\dagger\hat\psi_z} + \mean{\hat\psi^\dagger_{z'}\hat\psi_{z'}}} \covar{\hat\psi_z}{\hat\psi_{z'}} \\
\im\,\dot{\mathcal N} &\covar{\hat\psi_z^\dagger}{\hat\psi_{z'}}
= -\mean{\hat\psi_z^{\dagger2}} \covar{\hat\psi_z}{\hat\psi_{z'}} + \mean{\hat\psi_{z'}^2} \covar{\hat\psi_z^\dagger}{\hat\psi^\dagger_{z'}} \nonumber\\
&\qquad{}- 2\paren[\big]{\mean{\hat\psi_z^\dagger \hat\psi_z} - \mean{\hat\psi^\dagger_{z'} \hat\psi_{z'}}} \covar{\hat\psi_z^\dagger}{\hat\psi_{z'}},
\end{align}
\end{subequations}
where we use the shorthand notations $\mean{\hat\psi_z^2} = \mean{\hat\psi_z}^2 + \var[\big]{\hat\psi_z}$, $\mean{\hat\psi^\dagger_z \hat\psi_z} = \abs{\mean{\hat\psi_z}}^2 + \covar{\hat\psi^\dagger_z}{\hat\psi_z}$, and $\covar{\hat\psi_z}{\hat\psi^\dagger_{z'}} = \covar{\hat\psi^\dagger_{z'}}{\hat\psi_z} + \delta(z-z')$.
(Note that the latter Dirac delta function is converted into a Kronecker delta upon discretizing of the continuum field following Appendix~\ref{sec:discretization}.)

To summarize, the GSSF equations of motion describing the propagation of both the mean field and the Gaussian quantum noise in $\chi^{(3)}$ waveguides are given by \eqref{eq:chi3-gaussian-mean-linear}, \eqref{eq:chi3-gaussian-covars-linear}, \eqref{eq:chi3-gaussian-mean-nonlin}, and \eqref{eq:chi3-gaussian-covars-nonlin}.

\subsection{Example: Soliton noise dynamics}

As a first demonstration, we apply GSSF to study propagation of a canonical Kerr soliton in a $\chi^{(3)}$ nonlinear waveguide.
Classically, the Kerr soliton is a perfectly stable waveform arising from the balance of linear dispersion with nonlinear self-phase modulation, and quantum noise around this classical solution, in the form of so-called ``Kerr squeezing'', has been extensively studied in quantum optics~\cite{Carter1987,Drummond1987,Guidry2022,Guidry2023}.
Conventionally, such studies use a linearized treatment~\cite{Hosaka2016, Haus1990, Helt2020} which, as discussed in Sec.~\ref{sec:single-mode}, presupposes a separation of energy scales between dynamics of the mean field and the quantum noise: The former occurs very quickly and is first solved using classical SSF, while the latter is treated as simple linear perturbations that follow the classical solution.
As we show, however, nonlinear dynamics captured by our GSSF method can have a qualitative impact on Kerr squeezing in the regime of small soliton amplitude (i.e., under stronger optical nonlinearities).

The Kerr soliton can be canonically treated using the $\chi^{(3)}$ nonlinear waveguide propagation model \eqref{eq:chi3-hamiltonian-continuum}, and we assume a quadratic dispersion where $\Omega(k) = \frac12 \omega'' k^2$.
In this case, the mean-field limit of \eqref{eq:nlse-quantum} (i.e., the classical NLSE) supports the well known sech-soliton solution~\cite{Agrawal2019}
\begin{subequations} \label{eq:sech-soliton}
\begin{align}
\psi_z^{\text{(sech)}} = \sqrt{\frac{\bar{n}}{2z_{\bar n}}} \exp\paren*{\frac{\im\pi t}{4t_{\bar n}}} \sech\paren*{\frac{z}{z_{\bar n}}},
\end{align}
where $\bar{n}$ is the mean photon number of the soliton, and
\begin{align}
t_{\bar n} &= \frac{2\pi\omega''}{g^2\bar{n}^2}, &
z_{\bar n} &=-\frac{2\omega''}{g\bar{n}}
\end{align}
\end{subequations}
are the characteristic soliton period and pulse width, respectively.
Note that we assume the regime of modulation instability $g\omega''<0$ for the soliton solution to exist.
For the purposes of this example, we initialize the pulse as a coherent state described by \eqref{eq:sech-soliton}.
Note that after scaling $t$ and $z$ by $t_{\bar n}$ and $z_{\bar n}$, respectively, the only free parameter is the mean photon number $\bar n$, which effectively captures the ``quantumness'' of the system.

A convenient way to analyze the quantum noise dynamics is to calculate the \emph{squeezing supermodes} of the field and their respective squeezing levels~\cite{Wasilewski2006, Lvovsky2007}.
Physically, for a pure Gaussian state, the squeezing supermodes correspond to a set of orthogonal pulse waveforms which independently experience quadrature squeezing.
In most cases, only a few dominant supermodes (corresponding to low-order waveforms) experience significant squeezing, thus providing an efficient description of the multimode squeezing and entanglement in the pulse.
Appendix~\ref{sec:supermode} summarizes the procedure we use to calculate these squeezing supermodes and their squeezing levels, using the covariance matrix produced by a numerical method like GSSF.
In general, the waveforms of the squeezing supermodes can dynamically \emph{change} throughout propagation~\cite{Gouzien2020}, and their shapes are also independent of (though usually influenced by) the shape of the mean-field waveform.
This can happen even when the mean field is classically stable (as is the case for the soliton solution \eqref{eq:sech-soliton}), and the transfer of photons out of the mean field and into the various squeezing supermodes effectively constitutes a quantum-noise-induced \emph{destabilization} of the stable classical solution.

\begin{figure}[bth]
\centering
\includegraphics[width=0.46\textwidth]{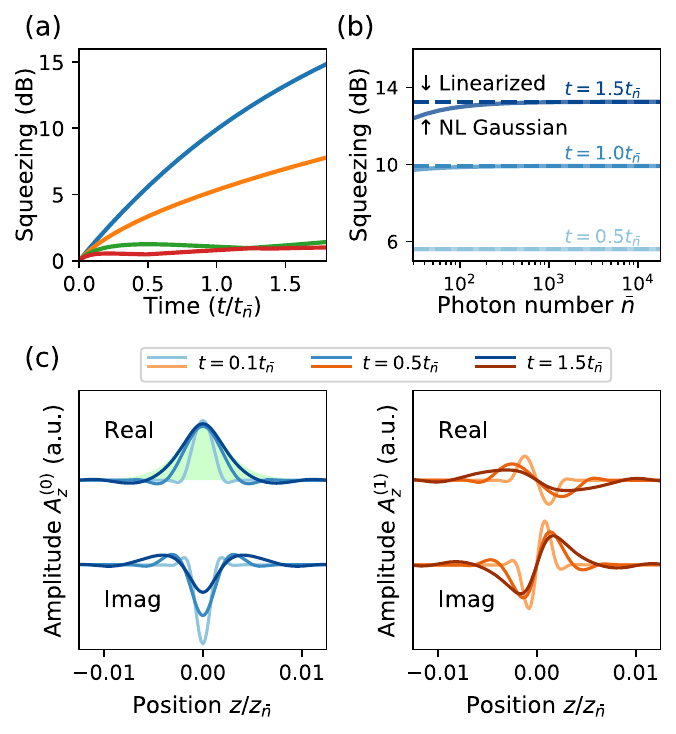}
\caption{
Quantum noise dynamics in $\chi^{(3)}$ propagation of a canonical sech soliton, instantiated as a coherent state given by \eqref{eq:sech-soliton}.
(a) Squeezing level of the four most major squeezing supermodes, simulated using GSSF for a mean photon number $\bar{n} = \num{1000}$.
(c) Evolution of the waveforms $A^{(0)}_z$ and $A^{(1)}_z$ that describe the two most major supermodes appearing in (a).
To contrast the squeezing supermodes against the classical solitonic waveform, we have multiplied $A^{(0)}_z$ and $A^{(1)}_z$ by $\exp\paren{-\im\pi t/4t_{\bar n}}$; the green shaded envelope shows the classical solitonic waveform $\propto \sech(z/z_{\bar n})$ for reference.
(b) The squeezing level of the most major squeezing supermode $A^{(0)}_z$ at various times, as a function of the mean photon number $\bar n$.
The prediction of the linearized treatment (which is independent of $\bar n$) is shown for comparison.
} \label{fig:soliton}
\end{figure}

Figure~\ref{fig:soliton} shows the quantum noise dynamics calculated by GSSF for a pulse initialized as a coherent-state soliton according to \eqref{eq:sech-soliton}.
Classically, the propagation dynamics are nearly trivial, with the classical waveform experiencing only a phase rotation $\e{\im\pi t/4t_{\bar n}}$ as given by \eqref{eq:sech-soliton}.
However, as shown in Fig.~\ref{fig:soliton}, there is continuous growth of squeezing in the pulse, occurring primarily in two squeezing supermodes, which we denote $A^{(0)}_z$ and $A^{(1)}_z$.
The waveforms describing these squeezing supermodes are shown at various propagation times in Fig.~\ref{fig:soliton}(c), from which we see that they clearly have significant transient behavior.
In particular, it is only after some propagation time ($t \gtrsim 1.5 t_{\bar n}$) that the real part of $A^{(0)}_z$ approaches that of the classical sech waveform.
However, even then there is a significant imaginary component (indicating a nonuniform phase shift from the classical envelope), as well as a significant amount of squeezing in the higher-order supermode $A^{(1)}_z$, which can be interpreted as timing jitter of the pulse due to quantum fluctuations~\cite{Haus1990}.
For a mean photon number of $\bar n = \num{1000}$ at which these simulations are done (and more generally in the semiclassical limit $\bar n \rightarrow \infty$), we note that many of these findings are in qualitative agreement with previous studies based on linearized treatments~\cite{Haus1990, Hosaka2016}.
At the same time, Fig.~\ref{fig:soliton}(b) shows that when we decrease $\bar n$ to ${}\sim \num{30}$, deviations appear between the linearized treatment and our nonlinear GSSF model, e.g., in the squeezing level.
Intuitively, this ``nonlinear saturation'' of the squeezing arises because the mean field becomes depleted to provide energy towards (anti)squeezing, thus limiting the effective gain available for further amplification of quantum fluctuations.

\section{Quantum noise propagation in a chi(2) waveguide} \label{sec:chi2-formalism}

Although the previous section treated the case of a $\chi^{(3)}$ waveguide, it should be clear that the moment-expansion and split-operator techniques can be readily generalized to other settings and optical nonlinearities as well.
Recently, $\chi^{(2)}$ waveguides in particular have shown experimental promise in being able to reach levels of optical nonlinearities where the nonlinear dynamics of quantum noise may become important. In this section, we apply the GSSF formalism to simulate $\chi^{(2)}$ nonlinear pulse propagation with conditions and parameters that are demonstrated recently, and we show that one could indeed observe strongly nonclassical and multimode photon dynamics in such experiments.

Due to the nature of the three-wave interactions characteristic to $\chi^{(2)}$ systems, it is often useful (though not required) to distinguish between fundamental- and second-harmonic bands in the spectrum of interacting modes.
In this two-envelope model, we introduce two fields $\phi_z$ and $\psi_z$ for the fundamental and second harmonic bands (FH and SH), respectively; as before, we assume $\sbrak[\big]{\hat\psi_z,\hat\psi_{z'}^\dagger} = \sbrak[\big]{\hat\phi_z,\hat\phi_{z'}^\dagger} = \delta(z-z')$ and define Fourier duals $\hat\Psi_k \coloneqq \int \dif z\, \e{-\im kz} \hat\psi_z$ (similarly for $\Phi_k$).
However, we also assume $\sbrak[\big]{\hat\psi_z,\hat\phi_{z'}^\dagger} = \sbrak[\big]{\hat\Psi_k,\hat\Phi_{k'}^\dagger} = 0$, i.e., that photons from the two bands are in principle distinguishable from one another (e.g., due to having different polarization or carrier-envelope phase).
Then, a suitable continuum Hamiltonian for this two-envelope model is
\begin{subequations}
 \label{eq:chi2-hamiltonian-continuum}
\begin{align}
\hat H_{\chi^{(2)}} &\coloneqq \hat H_\text{d3wm} + \hat H_\text{lin}, \quad\text{where} \\
\hat H_\text{d3wm} &\coloneqq \frac{\hbar}{2} \int \dif z \, \epsilon \paren*{\im\hat\psi_z \hat\phi_z^{\dagger2} - \im\hat\psi_z^\dagger \hat\phi_z^2} \\
\quad
\hat H_\text{lin} &\coloneqq \hbar \int \frac{\dif k}{2\pi} \paren*{\Omega_1(k) \hat\Phi_k^\dagger \hat\Phi_k + \Omega_2(k) \hat\Psi_k^\dagger \hat\Psi_k}.
\end{align}
\end{subequations}
Here, $\epsilon$ is a coupling rate related to the nonlinearity, while the FH dispersion is taken around the fundamental carrier wavevector $k_0$ with $\Omega_1(k) \coloneqq \omega(k_0+k) - \sbrak[\big]{\omega(k_0) + k\omega'(k_0)}$ and the SH dispersion is taken around $2k_0$ with $\Omega_2(k) \coloneqq \omega(2k_0+k) - \sbrak[\big]{2\omega(k_0) + k\omega'(k_0)}$.
That is, $\hat\Phi_k$ is, as usual, taken to be in a frame rotating at $\omega(k_0) + k\omega'(k_0)$, but $\hat\Psi_k$ rotates in an FH-derived frame at $2\omega(k_0) + k\omega'(k_0)$ and $\psi_z$ acts on relative positions $z$ which copropagate at the same speed $\omega'(k_0)$ as the FH fields $\phi_z$.
Note that with this convention, both phase and group-velocity mismatch are captured by $\Omega_2(k)$.
We refer to this two-envelope $\chi^{(2)}$ model as a (quasi)degenerate three-wave-mixing (3WM) model in analogy to degenerate 3WM in continuous-wave $\chi^{(2)}$ systems, where second-harmonic and half-harmonic generation are the dominant processes; in a multimode system with broadband phase matching, the 3WM is not strictly degenerate due to the energy difference between signal and idler within the FH band.

The Heisenberg equation of motions generated by $\hat H_{\chi^{(2)}}$ are
\begin{subequations}
\begin{align}
\partial_t \hat\phi_z &= \epsilon \hat\psi_z \hat\phi_z^\dagger - \im \Omega_1(-\im\partial_z) \hat\phi_z, \\
\partial_t \hat\psi_z &= -\frac{\epsilon}{2} \hat\phi_z^2 - \im \Omega_2(-\im\partial_z) \hat\psi_z.
\end{align}
\end{subequations}
The mean-field version of these equations, obtained by formally replacing $\hat\psi_z$ with $\psi(z)$ and $\hat\phi_z$ with $\phi(z)$ describing the classical SH and FH waveforms, respectively, are precisely the classical coupled-wave equations for $\chi^{(2)}$ waveguide propagation with linear dispersion.

To derive the GSSF model for this two-envelope $\chi^{(2)}$ model, we clearly need to track six covariances rather than two.
This aside, however, the entire procedure remains the same as in Sec.~\ref{sec:chi3-formalism}.
We begin with the split-operator quantum equations of motion
\begin{subequations}
\begin{align}
\dot{\mathcal N} \hat\phi_z &= \frac{\im}{\hbar\epsilon} \sbrak{\hat H_\text{d3wm}, \hat\phi_z} = \hat\psi_z \hat\phi_z^\dagger, \\
\dot{\mathcal N} \hat\psi_z &= \frac{\im}{\hbar\epsilon} \sbrak{\hat H_\text{d3wm}, \hat\psi_z} = -\frac 1 2 \hat\phi_z^2, \\
\dif\mathcal D \hat\Phi_k &= \frac{\im}{\hbar} \sbrak{\hat H_\text{lin}, \hat\Phi_k}\,\dif t = -\im\Omega_1(k) \hat\Phi_k \,\dif t, \\
\dif\mathcal D \hat\Psi_k &= \frac{\im}{\hbar} \sbrak{\hat H_\text{lin}, \hat\Psi_k}\,\dif t = -\im\Omega_2(k) \hat\Psi_k \,\dif t.
\end{align}
\end{subequations}
As usual, the dispersive step due to $\hat H_\text{lin}$ is linear so we straightforwardly have
\begin{subequations} \label{eq:chi2-gaussian-mean-linear}
\begin{align}
\im\,\dif\mathcal D \mean{\hat\Phi_k} &= \Omega_1(k)\mean{\hat\Phi_k} \, \dif t \\
\im\,\dif\mathcal D \mean{\hat\Psi_k} &= \Omega_2(k)\mean{\hat\Psi_k} \, \dif t
\end{align}
\end{subequations}
for the means, and, for the covariances,
\begin{subequations} \label{eq:chi2-gaussian-covars-linear}
\begin{align}
\im\,\dif\mathcal D\covar{\hat\Phi_k}{\hat\Phi_{k'}} &= \paren[\big]{\Omega_1(k') + \Omega_1(k)} \covar{\hat\Phi_k}{\hat\Phi_{k'}} \,\dif t \\
\im\,\dif\mathcal D\covar{\hat\Phi_k^\dagger}{\hat\Phi_{k'}} &= \paren[\big]{\Omega_1(k') - \Omega_1(k)} \covar{\hat\Phi_k^\dagger}{\hat\Phi_{k'}} \,\dif t \\
\im\,\dif\mathcal D\covar{\hat\Psi_k}{\hat\Psi_{k'}} &= \paren[\big]{\Omega_2(k') + \Omega_2(k)} \covar{\hat\Psi_k}{\hat\Psi_{k'}} \,\dif t \\
\im\,\dif\mathcal D\covar{\hat\Psi_k^\dagger}{\hat\Psi_{k'}} &= \paren[\big]{\Omega_2(k') - \Omega_2(k)} \covar{\hat\Psi_k^\dagger}{\hat\Psi_{k'}} \,\dif t \\
\im\,\dif\mathcal D\covar{\hat\Phi_k}{\hat\Psi_{k'}} &= \paren[\big]{\Omega_2(k') + \Omega_1(k)} \covar{\hat\Phi_k}{\hat\Psi_{k'}} \,\dif t \\
\im\,\dif\mathcal D\covar{\hat\Phi_k^\dagger}{\hat\Psi_{k'}} &= \paren[\big]{\Omega_2(k') - \Omega_1(k)} \covar{\hat\Phi_k^\dagger}{\hat\Psi_{k'}} \,\dif t.
\end{align}
\end{subequations}

The nonlinear step is as usual more involved, and using the same moment expansion methods, we can derive
\begin{subequations} \label{eq:chi2-gaussian-mean-nonlin}
\begin{align}
\dot{\mathcal N} \mean{\hat\phi_z} &= \mean{\hat\psi_z} \mean{\hat\phi_z^\dagger} + \covar{\hat\phi_z^\dagger}{\hat\psi_z} \\
\dot{\mathcal N} \mean{\hat\psi_z} &= -\frac{1}{2} \paren*{\mean{\hat\phi_z}^2 + \var{\hat\phi_z}}
\end{align}
\end{subequations}
for the means, and for the covariances,
\begin{subequations} \label{eq:chi2-gaussian-covars-nonlin}
\begin{align}
\dot{\mathcal N}\covar{\hat\phi_z}{\hat\phi_{z'}} &= \mean{\hat\phi_z^\dagger} \covar{\hat\phi_{z'}}{\hat\psi_z} + \mean{\hat\psi_z} \covar{\hat\phi_z^\dagger}{\hat\phi_{z'}} \\
&\qquad{}+ \mean{\hat\phi_{z'}^\dagger} \covar{\hat\phi_z}{\hat\psi_{z'}} + \mean{\hat\psi_{z'}} \covar{\hat\phi_z}{\hat\phi_{z'}^\dagger} \nonumber\\
\dot{\mathcal N}\covar{\hat\phi_z^\dagger}{\hat\phi_{z'}} &= \mean{\hat\phi_z} \covar{\hat\phi_{z'}}{\hat\psi_z^\dagger} + \mean{\hat\psi_z^\dagger} \covar{\hat\phi_z}{\hat\phi_{z'}} \\
&\qquad{}+ \mean{\hat\phi_{z'}^\dagger}\covar{\hat\phi_z^\dagger}{\hat\psi_{z'}} + \mean{\hat\psi_{z'}} \covar{\hat\phi_z^\dagger}{\hat\phi_{z'}^\dagger} \nonumber\\
\dot{\mathcal N}\covar{\hat\psi_z}{\hat\psi_{z'}} &= -\mean{\hat\phi_z}\covar{\hat\phi_z}{\hat\psi_{z'}} - \mean{\hat\phi_{z'}}\covar{\hat\phi_{z'}}{\hat\psi_z} \\
\dot{\mathcal N}\covar{\hat\psi_z^\dagger}{\hat\psi_{z'}} &= -\mean{\hat\phi_z^\dagger}\covar{\hat\phi_z^\dagger}{\hat\psi_{z'}} - \mean{\hat\phi_{z'}}\covar{\hat\phi_{z'}}{\hat\psi_z^\dagger} \\
\dot{\mathcal N}\covar{\hat\phi_z}{\hat\psi_{z'}} &= \mean{\hat\phi_z^\dagger}\covar{\hat\psi_z}{\hat\psi_{z'}} + \mean{\hat\psi_z}\covar{\hat\phi_z^\dagger}{\hat\psi_{z'}} \nonumber\\
&\qquad{}- \mean{\hat\phi_{z'}}\covar{\hat\phi_z}{\hat\phi_{z'}} \\
\dot{\mathcal N}\covar{\hat\phi_z^\dagger}{\hat\psi_{z'}} &= \mean{\hat\phi_z}\covar{\hat\psi_z^\dagger}{\hat\psi_{z'}} + \mean{\hat\psi_z^\dagger}\covar{\hat\phi_z}{\hat\psi_{z'}} \nonumber\\
&\qquad{}- \mean{\hat\phi_{z'}}\covar{\hat\phi_z^\dagger}{\hat\phi_{z'}}.
\end{align}
\end{subequations}

To summarize, the GSSF equations of motion for (quasi)degenerate three-wave-mixing in $\chi^{(2)}$ waveguides are given by \eqref{eq:chi2-gaussian-mean-linear}, \eqref{eq:chi2-gaussian-covars-linear}, \eqref{eq:chi2-gaussian-mean-nonlin}, and \eqref{eq:chi2-gaussian-covars-nonlin}.
It is also worth noting that, in the single-mode scenario, these nonlinear moment equations are consistent with the ones derived in Ref.~\cite{Huang2022} for studying single-mode $\chi^{(2)}$ interactions.

\subsection{Example: Pump depletion in pulsed squeezing}

The most successful schemes to date for generating squeezed light, especially for use as resource states in quantum metrology and continuous-variable quantum information processing, rely on phase-sensitive (degenerate) optical parametric amplification in materials with $\chi^{(2)}$ nonlinearities, in which SH pump induces quadrature squeezing on FH signal.
In the absence of a signal seed, this process produces a squeezed vacuum state via parametric deamplification of vacuum noise along one quadrature.
Conventionally, such squeezing experiments utilize a highly excited coherent-state pump in a weakly nonlinear $\chi^{(2)}$ crystal to generate vacuum squeezing with low conversion efficiency.
In this low-efficiency limit, the process is well described by an undepleted pump approximation in which the pump is in a static coherent state, i.e., an interaction Hamiltonian of the form $\hat\Psi_k \hat\Phi_{k'}^\dagger \hat\Phi_{k-k'}^\dagger + \text{H.c.} \approx \mean{\hat\Psi_k(0)} \hat\Phi_{k'}^\dagger \hat\Phi_{k-k'}^\dagger + \text{H.c.}$, leading to multimode but purely linear squeezing dynamics for the signal.
These dynamics can be integrated to obtain a linearized estimate of the signal covariance matrix.

Recently, however, dispersion engineering in tightly confining TFLN waveguides has enabled a significant increase in the effective nonlinearity of $\chi^{(2)}$ parametric interactions, challenging the conventional undepleted pump approximation.
For example, Refs.~\cite{Jankowski2022,Ledezma2022} experimentally demonstrated waveguides that can support $\approx\SI{70}{dB}$ of broadband parametric gain with only a \SI{4}{\pico\joule} pump pulse.
Heuristically, \SI{70}{dB} of antisqueezing at \SI{2}{\micro\meter} corresponds to ${}\approx\SI{0.5}{\pico\joule}$ of parametric fluorescence per pulse, suggesting that state-of-the-art devices can exhibit ${}>\SI{10}{\percent}$ pump depletion solely through the amplification of vacuum fluctuations.
The regime where parametric fluorescence is sufficiently bright to deplete the pump is commonly referred to as optical parametric generation (OPG), and the effects of pump depletion on squeezing have previously been studied in the single-mode case~\cite{Degenfeld-Schonburg2015,Veits1995}.
Here, we employ GSSF to analyze the dynamics of saturated OPG in the ultrafast domain, looking in particular at the intrinsically multimode entanglement structure of the output parametric fluorescence.

First, however, it is worth noting that for vacuum-seeded OPG, the nonlinear equations of motion \eqref{eq:chi2-gaussian-mean-nonlin} and \eqref{eq:chi2-gaussian-covars-nonlin} take a particularly simple form.
Since the initial input signal field is vacuum, $\mean{\hat\phi_z} = 0$ at $t = 0$.
Furthermore, since the initial input pump field is a coherent state, the pump and signal are initially uncorrelated, so $\covar{\hat\phi_z}{\hat\psi_{z'}} = \covar{\hat\phi_z^\dagger}{\hat\psi_{z'}^\dagger} = 0$ at $t = 0$.
Then by inspection of \eqref{eq:chi2-gaussian-mean-nonlin} and \eqref{eq:chi2-gaussian-covars-nonlin}, we see that, for all time,
\begin{align}
\mean{\hat\phi_z} &= \covar{\hat\psi_z}{\psi_{z'}} = \covar{\hat\psi_z^\dagger}{\psi_{z'}} \nonumber\\
&= \covar{\hat\phi_z}{\psi_{z'}} = \covar{\hat\phi_z^\dagger}{\hat\psi_{z'}} = 0.
\end{align}
In fact, the only non-trivial dynamics are in mean of the pump and the covariances of the signal, given by
\begin{subequations} \label{eq:chi2-squeezing-eoms}
\begin{equation}
\dot{\mathcal N} \mean{\hat\psi_z} = -\frac 1 2 \var{\hat\phi_z} \end{equation}
for the mean, and, for the covariances,
\begin{align}
\dot{\mathcal N} \covar{\hat\phi_z}{\hat\phi_{z'}} &= \mean{\hat\psi_z}\covar{\hat\phi_z^\dagger}{\hat\phi_{z'}} + \mean{\hat\psi_{z'}}\covar{\hat\phi_z}{\hat\phi_{z'}^\dagger} \\
\dot{\mathcal N} \covar{\hat\phi_z^\dagger}{\hat\phi_{z'}} &= \mean{\hat\psi_z^\dagger}\covar{\hat\phi_z}{\hat\phi_{z'}} + \mean{\hat\psi_{z'}}\covar{\hat\phi_z^\dagger}{\hat\phi_{z'}^\dagger}.
\end{align}
\end{subequations}
We see that even in the Gaussian-state approximation, there are nonlinear dynamics in which the pump mean experiences depletion due to the generation of signal photon pairs.
The pump also remains in a coherent state and is unentangled with the signal, which need not hold true in more exotic non-Gaussian settings~\cite{Yanagimoto2021-non-gaussian}.

\begin{figure}[t]
\centering
\includegraphics[width=\columnwidth]{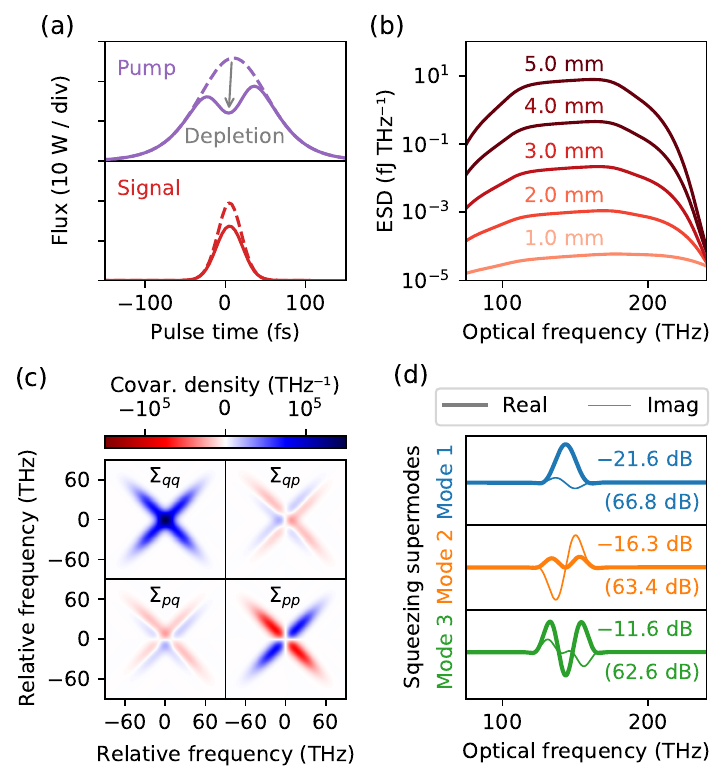}
\caption{
Squeezing in saturated optical parametric generation by a nonlinear $\chi^{(2)}$ waveguide.
(a) Pulse envelopes $\abs{\mean{\hat\psi_z}}^2$ (top) and $\covar{\hat\phi_z^\dagger}{\hat\phi_z}$ (bottom) for the coherent pump and signal fluorescence, respectively, expressed in photon flux as a function of pulse temporal coordinate.
Dashed lines indicate predictions of a linearized model assuming undepleted pump.
(b) Spectrum of signal fluorescence $\covar{\hat\Phi_k^\dagger}{\hat\Phi_k}$, expressed in energy spectral density (ESD) as a function of optical frequency, shown at various propagation lengths.
(c) Covariance matrix $\Sigma$ of the signal (see Appendix~\ref{sec:supermode}), indexed by relative frequency around the signal carrier.
Units are chosen such that the integral of the diagonal of $\Sigma$ minus that of the vacuum produces the total number of fluorescence photons.
(d) Spectra of the three most major squeezing supermodes (thick line: real part; thin line: imaginary part) shown together with their respective levels of quadrature squeezing and antisqueezing (latter in parentheses).
See Table~\ref{tab:chi2-parameters} for detailed parameters for this simulation.
} \label{fig:chi2-squeezing}
\end{figure}

Figure~\ref{fig:chi2-squeezing} shows a GSSF simulation of OPG in a waveguide with parameters similar to that of Ref.~\cite{Jankowski2022} (see Table~\ref{tab:chi2-parameters}), using the simplified equations \eqref{eq:chi2-squeezing-eoms}, with minor modifications to account for linear loss as discussed in Appendix~\ref{sec:dissipation}.
As expected, Fig.~\ref{fig:chi2-squeezing}(a) shows that the pump experiences a significant amount of depletion, with a dip generated as the signal fluorescence grows and walks off from the center.
This process amounts to a nonlinear saturation of the parametric gain even under vacuum input.
The waveforms predicted by GSSF differ significantly from those in the linearized model, which we plot as corresponding dashed lines:
In the latter, the pump amplitude experiences only dispersion and loss, which causes the model to overestimate the signal fluorescence due to the absence of nonlinear saturation.
In this simulation, we find $\sim\SI{0.6}{\pico\joule}$ of pump depletion per pulse, in accordance with energy conservation and in rough agreement with measurements reported in Refs.~\cite{Jankowski2022, Ledezma2022}.
Figure~\ref{fig:chi2-squeezing}(b) shows that the spectrum of the signal fluorescence is in qualitative agreement with experiments as well.

Our numerical results also reveal the quantum correlation structure of the squeezed light produced by OPG, which to our knowledge have yet to be fully explored experimentally.
Figure~\ref{fig:chi2-squeezing}(c) shows the covariance matrix of the signal in the frequency domain, which fully characterizes the Gaussian quantum state of the signal pulse.
Because OPG produces squeezing and antisqueezing predominantly along the quadratures of the field (see Appendix~\ref{sec:supermode}), we focus on the covariance matrix written in the quadrature basis $(\hat q_z, \hat p_z)$.
The spectral correlations in the covariance matrix indicate that the signal field occupies a multimode squeezed state with significant levels of entanglement among many spectral-temporal components.
It is worth noting that such correlations are lost when observing only the fluorescence spectrum, viz., Fig.~\ref{fig:chi2-squeezing}(b), and more sophisticated techniques in quantum state tomography of ultrafast pulses are needed to probe the covariance structure in greater detail~\cite{Nehra2022}.

To further understand the entanglement structure, we can also utilize a supermode decomposition of the covariance matrix (as discussed in Appendix~\ref{sec:supermode}) to obtain the dominant squeezing supermodes in the signal pulse, which we show in Fig.~\ref{fig:chi2-squeezing}(d).
Whereas any given spatial bin or narrowband component of the signal is highly entangled with the rest of the field, the squeezing supermodes comprise a superposition of many narrowband components, chosen in such a way that they are minimally correlated (i.e., unentangled or separable) with one another.
In other words, measurements selectively probing these squeezing supermodes (i.e., via a pulse-shaped local oscillator in optical homodyne) are needed to fully decompose the Gaussian state into its independently squeezed components.
We find that the dominant supermode, as expected, has a spectrum that is mostly determined by the pump spectrum, with a slight variation in spectral phase imposed by dispersion (in particular, group velocity mismatch).
This supermode experiences nearly \SI{67}{\deci\bel} of antisqueezing, in agreement with empirical estimates of the parametric gain in Refs.~\cite{Jankowski2022,Ledezma2022}.
We also see there are at least two other supermodes all experiencing ${}>\SI{60}{\deci\bel}$ of gain, which is expected as the device is not specifically engineered to exclusively provide gain in a single supermode; advanced engineering of OPG devices may enable more efficient channeling of pump energy into selectively squeezing specific supermode patterns of interest.
Perhaps most interestingly, however, we observe that the dominant squeezing supermode retains up to \SI{20}{\deci\bel} of quadrature squeezing despite the fact that our simulations already take into account a propagation loss of \SI{30}{\deci\bel\per\meter} for the signal field.
Further work developing and deploying this potent control over the behavior of quantum noise (e.g. by reducing propagation losses and increasing outcoupling efficiencies) appears to be highly worthwhile for advancing the state of the art in quantum photonics.

\begin{table}[t]
\centering
\setlength{\extrarowheight}{2pt}
\begin{tabular}{l r r}
\hline\hline
Device Parameter & Figure 2 & Figure 3 \\
\hline
FH Wavelength & \SI{2090}{\nano\meter} & \SI{2090}{\nano\meter} \\
Normalized SHG Efficiency & \SI{10}{\per\watt\per\cm\squared} & \SI{10}{\per\watt\per\cm\squared} \\
Waveguide Length $L$ & \SI{5.0}{\milli\meter} & \SI{6.0}{\milli\meter} \\
Phase Mismatch & 0 & $-3\pi/L$ \\
Group Velocity Mismatch & \SI{2.0}{\femto\second\per\milli\meter} & \SI{10}{\femto\second\per\milli\meter} \\
FH Group Velocity Dispersion & \SI{10}{\femto\second\squared\per\milli\meter} & \SI{-15}{\femto\second\squared\per\milli\meter} \\
SH Group Velocity Dispersion & \SI{100}{\femto\second\squared\per\milli\meter} & \SI{100}{\femto\second\squared\per\milli\meter} \\
FH Third-order Dispersion & 0 & \SI{500}{\femto\second\cubed\per\milli\meter} \\
SH Third-order Dispersion & 0 & \SI{1000}{\femto\second\cubed\per\milli\meter} \\
FH Input Pulse Parameters & 0 & \SI{5.0}{\pico\joule}, \SI{50}{\femto\second} \\
SH Input Pulse Parameters & \SI{3.0}{\pico\joule}, \SI{100}{\femto\second} & 0 \\
\hline\hline
\end{tabular}
\caption{
Summary of parameters used for $\chi^{(2)}$ waveguide simulations in Figs.~\ref{fig:chi2-squeezing} and \ref{fig:chi2-supercontinuum}, based on Ref.~\cite{Jankowski2022} and Ref.~\cite{Jankowski2021}, respectively.
Input pump pulse parameters describes the pulse energy and full width at half-maximum of a $\sech^2$ envelope, taken to be in a coherent state.
Loss model assumes constant FH attenuation of \SI{30}{dB/m} rising to \SI{20}{dB/cm} after a cutoff at \SI{2900}{nm} and negligible attenuation in the SH.
} \label{tab:chi2-parameters}
\end{table}

\subsection{Example: Quantum noise in second-order supercontinuum generation} \label{sec:supercontinuum}

\begin{figure*}[h!t]
\centering
\includegraphics[width=2\columnwidth]{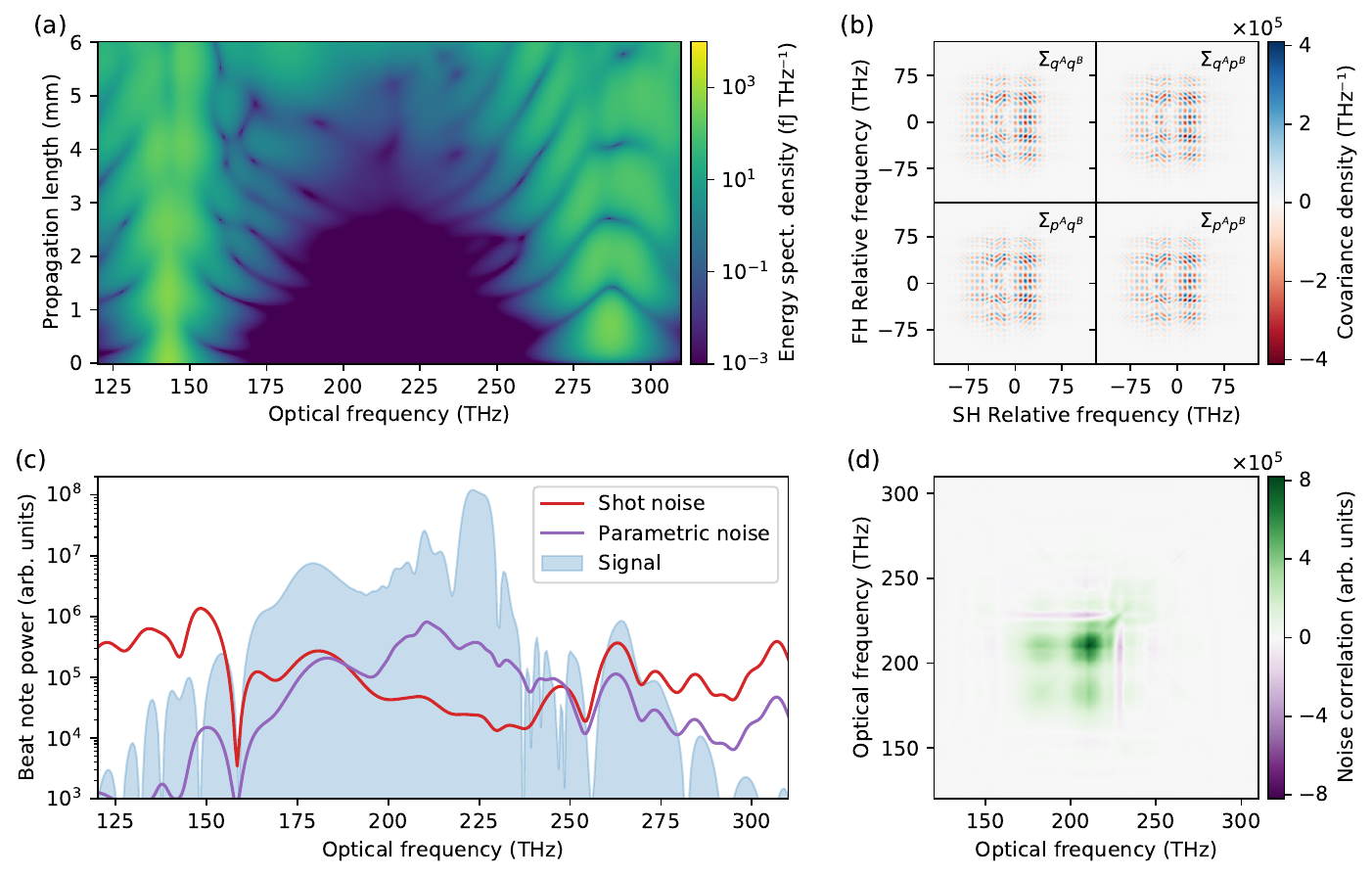}
\caption{
Quantum correlations in a $\chi^{(2)}$ supercontinuum and its fundamental noise contribution to CEO frequency detection.
(a) Energy spectral density $E(f)$ as a function of optical frequency, where $E(mf_\text{rep}) \coloneqq \mean{\hat A_m^\dagger\hat A_m} + \mean{\hat B_{q(m)}^\dagger\hat B_{q(m)}}$; c.f., Ref.~\cite{Jankowski2021}.
(b) Specific subblocks of the two-envelope covariance matrix $\Sigma$ (see Appendix~\ref{sec:supermode}) showing correlations between the $m$th FH comb line and the $q$th SH comb line, indexed by relative frequencies $mf_\text{rep}$ and $qf_\text{rep}$ around their respective carriers at $m_0f_\text{rep}$ and $2m_0 f_\text{rep}$.
Units are chosen such that the integral of the diagonal of $\Sigma$ minus that of the vacuum produces the total number of fluorescence photons.
(c) Photocurrent signal and noise power at $f_\text{ceo}$ generated by beating the $m$th FH comb line against the corresponding SH comb line at $q(m) = m-m_0$, plotted as a function of the optical frequency $f = mf_\text{rep}$.
The signal is given by $S^2(f)$ according to \eqref{eq:fceo-signal}, the shot noise is given by $S_\text{shot}(f)$ \eqref{eq:fceo-shot}, and the excess (parametric) noise is given by $S_\text{para}(f)$ \eqref{eq:fceo-parametric}, where we have assumed $\phi_\text{ceo} = \pi/3$ in \eqref{eq:fceo-variance}.
(d) Correlations between fluctuations in the $f_\text{ceo}$ beat note generated by different parts of the frequency comb, as captured by $N_1(f,f') + N_2(f,f')$ from \eqref{eq:fceo-variance}, also assuming $\phi_\text{ceo} = \pi/3$.
See Table~\ref{tab:chi2-parameters} for detailed parameters for this simulation.
} \label{fig:chi2-supercontinuum}
\end{figure*}

We conclude this section by applying our GSSF method to study quantum noise in supercontinuum generation (SCG).
It was recently reported in Refs.~\cite{Jankowski2020, Jankowski2021} that highly efficient, saturated second-harmonic generation in a dispersion-engineered TFLN waveguide with a modest amount of phase mismatch can dynamically produce such strong modulations on the FH and SH envelopes that their collective bandwidths span more than an octave in frequency.
Classical analysis reveals that a relatively simple model involving only coherent $\chi^{(2)}$ nonlinear interactions between the FH and SH envelopes are sufficient to explain most of the qualitative features in the supercontinuum spectrum~\cite{Jankowski2021}, making $\chi^{(2)}$ SCG an interesting example on which to test our method, even without accounting for auxiliary effects like stimulated Raman, etc., which can play important roles in $\chi^{(3)}$ SCG~\cite{Dudley2006}.
An important application of SCG is enabling detection of the carrier-envelope-offset (CEO) frequency, $f_\text{ceo}$, of a frequency comb~\cite{Jones2000,Helbing2003}, which is essential for building a stable clockwork for optical frequency metrology~\cite{Diddams2000,Holzwarth2000}.
Because both the FH and SH envelopes are produced in the broadening dynamics in $\chi^{(2)}$ SCG, the waveguide output can be directly heterodyned to generate a beat note at $f_\text{ceo}$ without the need for an additional frequency-doubling stage~\cite{Jankowski2020,Jankowski2021}.
For use in optical frequency metrology, however, it is important that we understand the fundamental and practical noise limits set by the SCG process~\cite{Corwin2003,Ames2003}.
Due to its parametric and coherent nature, the $\chi^{(2)}$ supercontinuum should exhibit quantum correlations and entanglement that can, in principle, increase (or even decrease) the noise in $f_\text{ceo}$ detection relative to a shot-noise assumption.
Here, we use the GSSF method to simulate the $\chi^{(2)}$ SCG dynamics, and we apply the quantum theory of heterodyne detection to the resulting Gaussian state in order to quantify the fundamental noise in the beat note signal set by such quantum fluctuations.

Figure~\ref{fig:chi2-supercontinuum} shows a GSSF simulation of a waveguide with parameters similar to that of Refs.~\cite{Jankowski2020, Jankowski2021} (see Table~\ref{tab:chi2-parameters}), using the full equations of motion for quasidegenerate three-wave-mixing.
The envelope spectral dynamics in Fig.~\ref{fig:chi2-supercontinuum}(a) shows good qualitative agreement with both classical simulations and experimental data~\cite{Jankowski2020,Jankowski2021}, with the formation of spectral overlap between FH and SH within a propagation length of \SI{3}{mm}.
Diving deeper into the quantum structure of the supercontinuum, however, Fig.~\ref{fig:chi2-supercontinuum}(b) shows the subblocks of the covariance matrix (see Appendix~\ref{sec:supermode}) which describe correlations between the FH and SH envelopes at the end of the waveguide.
We see finely patterened correlations with complex spectral structure imparted by the dynamics of the nonlinear SCG process.
Generically, such quantum correlations indicate the presence of multimode entanglement, and the squeezing supermodes of the total field consists of hybridized excitations of both FH and SH envelopes.

These correlations in the supercontinuum contribute to quantum-limited noise that is present when measuring the $f_\text{ceo}$ beat note through direct heterodyne detection, e.g., as done in Ref.~\cite{Jankowski2020}.
In Appendix~\ref{sec:heterodyne}, we use the standard quantum theory of optical heterodyne detection to derive both the signal and noise associated with measuring the $f_\text{ceo}$ beat note.
Because we are interested here in frequency combs rather than continuum fields, we introduce discrete FH modes $\hat A_m$ and SH modes $\hat B_q$ (see also Appendix~\ref{sec:discretization}) in place of $\hat\Phi_k$ and $\hat\Psi_k$, and we assume that the bare frequencies of these modes are $(m+m_0)f_\text{rep} + f_\text{ceo}$ and $(q+2m_0)f_\text{rep} + 2f_\text{ceo}$, respectively.
Here, $f_\text{rep}$ is the repetition rate of the comb, $m_0$ indexes the central comb line of the FH envelope, and $f_\text{ceo}$ is the CEO frequency of the FH envelope.
We find that the total steady-state photocurrent demodulated at $f_\text{ceo}$ is given by
\begin{subequations} \label{eq:fceo-signal}
\begin{equation}
    \mean{I_\text{h}} = {f_\text{rep}} \sum_{m} S(m f_\text{rep}) \\
\end{equation}
where the signal contribution from each comb line is
\begin{equation}
    S(m f_\text{rep}) \coloneqq \RE\mean{\hat A_m^\dagger\hat B_{q(m)}},
\end{equation}
\end{subequations}
where $q(m) \coloneqq m-m_0$ denotes the index of the SH comb mode $\hat B_{q(m)}$ whose beat with $\hat A_m$ contributes to the signal at $f_\text{ceo}$.
It is interesting to note that $\mean{\hat A_m^\dagger\hat B_{q(m)}} = \mean{\hat A_m^\dagger}\mean{\hat B_{q(m)}} + \covar{\hat A_m^\dagger}{\hat B_{q(m)}}$, i.e., there are contributions to the beat note not only from the mean field (first term) but also from the quantum correlations (second term) between the two envelopes.

The noise on the photocurrent signal is characterized by the total variance
\begin{subequations} \label{eq:fceo-variance}
\begin{align}
&\frac{\mean{\delta I_\text{h}^2}}{f_\text{rep}^2} = \sum_m N_0(m f_\text{rep})
+ \frac12 \sum_{m,m'} N_1(mf_\text{rep}, m'f_\text{rep}) \nonumber\\
&\qquad{}+ \frac12 \sum_{m,m'} N_2(mf_\text{rep}, m'f_\text{rep}) \paren*{1 + \sinc^2 \phi_\text{ceo}}.
\end{align}
where $\phi_\text{ceo} \coloneqq 2\pi f_\text{ceo}/f_\text{rep}$ is the relative CEO phase accumulated by successive pulses and
\begin{align}
N_0(m f_\text{rep}) &\coloneqq \mean{\hat A_m^\dagger \hat A_m} + \mean{\hat B_m^\dagger \hat B_m}, \\
N_1(mf_\text{rep}, m'f_\text{rep}) &\coloneqq \RE\mean[\big]{\hat A_m^\dagger \hat A_{m'}^\dagger \hat B_{q(m')} \hat B_{q(m)}} \\
&\quad{}- \RE\sbrak[\Big]{\mean[\big]{\hat A_m^\dagger \hat B_{q(m)}} \mean[\big]{\hat A_{m'}^\dagger \hat B_{q(m')}}}, \nonumber\\
N_2(mf_\text{rep}, m'f_\text{rep}) &\coloneqq \RE\mean[\big]{\hat A_m^\dagger \hat A_{m'} \hat B_{q(m')}^\dagger \hat B_{q(m)}} \\
&\quad{}- \RE\sbrak[\Big]{\mean[\big]{\hat A_m^\dagger \hat B_{q(m)}} \mean[\big]{\hat A_{m'} \hat B_{q(m')}^\dagger}}. \nonumber
\end{align}
\end{subequations}
This rather complicated expression (see Appendix~\ref{sec:heterodyne} for more details) arises because the beat-note fluctuations coming from two different frequency indices $m$ and $m'$ can, in principle, be (anti)-correlated, thus leading to an increase (decrease) in the total noise when summed together.
The uncorrelated fluctuations are captured by $N_0$ and the diagonal components of $N_1$ and $N_2$.
Note that for $N_0$, we can write, e.g., $\mean{\hat A_m^\dagger \hat A_m} = \abs{\mean{\hat A_m}}^2 + \covar{\hat A_m^\dagger}{\hat A_m}$; the first term is the standard shot noise of the mean field, while the second term is excess noise due to parametric fluorescence.
On the other hand, the correlated fluctuations $N_1$ and $N_2$ are related to fourth-order moments of the state; for a Gaussian state these fourth-order correlations can be reduced to second-order correlations and evaluated readily (see Appendix~\ref{sec:heterodyne}).

In Fig.~\ref{fig:chi2-supercontinuum}(c), we show the spectrum of the beat-note signal $S^2(f)$ \eqref{eq:fceo-signal} and the diagonal contributions from the noise \eqref{eq:fceo-variance} (physically corresponding to the use of a tunable narrowband optical filter in front of the detector).
We separate the noise into ``shot noise'' and ``parametric noise'' contributions according to
\begin{subequations}
\begin{align}
    N_\text{shot}(mf_\text{rep}) &\coloneqq \abs{\mean{\hat A_m}}^2 + \abs{\mean{\hat B_{q(m)}}}^2 \label{eq:fceo-shot} \\
    N_\text{para}(mf_\text{rep}) &\coloneqq \covar{\hat A_m^\dagger}{\hat A_m} + \covar{\hat B_{q(m)}^\dagger}{\hat B_{q(m)}} \label{eq:fceo-parametric} \\
    &\quad{}+ N_1(mf_\text{rep},mf_\text{rep}) + N_2(mf_\text{rep},mf_\text{rep}), \nonumber
\end{align}
\end{subequations}
where $N_\text{para}$ captures the excess noise (beyond shot noise) coming from the fluorescence and quantum correlations in the supercontinuum.
Perhaps surprisingly, we find that the parametric noise is comparable to---and in certain parts of the spectrum even in excess of---the shot noise predicted by the mean field.
Finally, to get a better sense for the off-diagonal correlations occurring in \eqref{eq:fceo-variance}, we show in Fig.~\ref{fig:chi2-supercontinuum}(d) the full correlation matrix $N_1(f,f') + N_2(f,f')$.
We see that there is indeed some degree of correlation between the beat note fluctuations coming from different parts of the frequency comb, suggesting we may be able to improve the signal-to-noise ratio via selective, multiband filtering of the supercontinuum prior to self-heterodyne detection.


\section{Conclusions}
In this work, we have developed a Gaussian split-step Fourier (GSSF) framework which integrates the formalism of Gaussian-state quantum optics with the nonlinear physics of ultrafast pulse propagation.
This GSSF method generalizes the classical SSF method to treat quantum fluctuations and correlations, up to second order, on an equal footing with mean-field nonlinear pulse dynamics.
Taking inspiration from state-of-the-art dispersion-engineered devices on thin-film lithium niobate, we have shown, through detailed case studies, how the GSSF method enables us to better understand both the operational principles and technological potential of photonic hardware near the quantum-classical transition.
For saturated optical parametric generation~\cite{Ledezma2022, Jankowski2022}, we have identified squeezing supermodes and their respective squeezing levels despite the presence of significant pump depletion, which puts this system beyond the scope of conventional linearized treatments of vacuum squeezing (i.e., via undepeleted-pump approximations).
For supercontinuum generation based on saturated second-harmonic generation~\cite{Jankowski2020, Jankowski2021}, we have used GSSF to resolve finely patterned spectral correlations inside the octave-spanning supercontinuum. We then leveraged standard quantum-optical theory to explicitly evaluate the quantum noise floor for $f-2f$ beat-note detection of the CEO frequency using this novel supercontinuum source, finding contributions beyond the shot-noise limit due to parametric fluorescence and frequency-domain entanglement.
These case studies demonstrate the effectiveness of the GSSF framework for analyzing and engineering ultrafast quantum photonic devices, and we expect in future work that even more sophisticated systems, from cavity-based frequency microcombs~\cite{Kippenberg2018} to nanophotonic synchronously-pumped optical parametric oscillators, can be straightforwardly treated as well.

Numerically, GSSF can directly leverage the remarkable efficiency of classical SSF methods, with each split-step requiring only $\mathcal O(M^2\log M)$ cost to update all $M^2$ Gaussian moments of an $M$-mode pulse.
It is also worth pointing out that GSSF generates all the Gaussian moments in a \emph{single} simulation, as opposed to mean-field Monte-Carlo techniques that require many trajectories to statistically resolve, e.g., small spectral features in the correlations.
We remark that even with a fairly na\"ive RK4IP implementation~\cite{Hult2007} of GSSF on an Ampere A100 GPU, the supercontinuum simulation of Sec.~\ref{sec:supercontinuum} requires ${}<\SI{5}{\second}$ using $M = 2^{10}$ points per envelope.

Finally, while this work has immediate practical relevance to current and near-term experiments operating in the semiclassical domain, the Gaussian-state framework, and GSSF by extension, is expected to remain indispensable well beyond the classical-quantum threshold.
For example, faithful descriptions of multimode squeezed states are essential for reliably generating the non-Gaussian resource states at the heart of the most mature photonic schemes for continuous-variable quantum computation~\cite{Walschaers2020,Bourassa2021}.
Even in the strong-coupling regime where non-Gaussian features emerge coherently, a Gaussian approximation to quantum dynamics provides vital information on how and where (i.e., in which supermodes) such non-Gaussian features appear, facilitating significantly more concise quantum state representations for pulse dynamics~\cite{Yanagimoto2021-non-gaussian}.
Thus, we expect this work to not only serve as a workhorse method for engineering near-term nonlinear ultrafast devices, but to also guide new conceptual and modeling paradigms for quantum photonics generally, by embracing rather than abstracting away the rich physics of ultrafast quantum dynamics.
\begin{appendix}

\begin{acknowledgments}
The authors are grateful to Logan G.\ Wright, Melissa A.\ Guidry, Daniil M.\ Lukin, Rajveer Nehra, and Alireza Marandi for helpful discussions.

The authors wish to thank NTT Research for their financial and technical support.
This work has been supported by the Army Research
Office under Grant No. W911NF-16-1-0086, and the National Science Foundation under awards CCF-1918549
and PHY-2011363.
\end{acknowledgments}

\section{Conservation of energy} \label{sec:conservation}

As discussed in Sec.~\ref{sec:single-mode} for the single-mode case, a distinguishing feature of our nonlinear Gaussian-state approximation compared with conventional linearized treatments is the conservation of photon(-polariton) number, i.e., energy, in the absence of linear losses.
In this section, we explicitly show how this conservation property arises in the multimode using the nonlinear Gaussian equations of motion for $\chi^{(3)}$ and $\chi^{(2)}$ waveguide propagation developed in Secs.~\ref{sec:chi3-formalism} and \ref{sec:chi2-formalism}.

First, let us consider the case of $\chi^{(3)}$ nonlinear propagation, where the total energy is given by the total photon number
\begin{equation}
\bar n \coloneqq \int\dif z\, \mean{\hat\psi_z^\dagger\hat\psi_z} = \int\frac{\dif k}{2\pi}\, \mean{\hat\Psi_k^\dagger\hat\Psi_k},
\end{equation}
where $\mean{\hat\psi_z^\dagger\hat\psi_z} = \mean{\hat\psi_z^\dagger}\mean{\hat\psi_z} + \covar{\hat\psi_z^\dagger}{\hat\psi_z}$, i.e., the sum of both classical (mean-field) and quantum-noise (diagonal covariance) contributions.
For linear evolution under $\hat{H}_\text{lin}$ in \eqref{eq:chi3-hamiltonian-continuum}, we straightforwardly have
\begin{subequations}
\begin{align}
\dot{\mathcal D} \paren{\mean{\hat\Psi_k^\dagger}\mean{\hat\Psi_k}} &= \dot{\mathcal D} \covar{\hat\Psi_k^\dagger}{\hat\Psi_k} = 0 \\
\Rightarrow \dot{\mathcal D} \bar n &= 0,
\end{align}
\end{subequations}
since $\Omega^*(k) = \Omega(k)$ (i.e., we only have dispersion) in the absence of loss.
On the other hand, for nonlinear evolution under $\hat H_\text{4wm}$, we can calculate
\begin{subequations}
\begin{align}
\im\,\dot{\mathcal N} \paren{\mean{\hat\psi_z^\dagger} \mean{\hat\psi_z}} &= \im \mean{\hat\psi_z^\dagger} \dot{\mathcal N}\mean{\hat\psi_z} + \im \mean{\hat\psi_z} \dot{\mathcal N}\mean{\hat\psi_z^\dagger} \nonumber\\
&= \mean{\hat\psi_z^\dagger}^2 \var{\hat\psi_z} - \mean{\hat\psi_z}^2 \mean{\delta\hat\psi_z^{\dagger2}} \\
\im\,\dot{\mathcal N} \covar{\hat\psi_z^\dagger}{\hat\psi_z} &= -\mean{\hat\psi_z^\dagger}^2 \var{\hat\psi_z} + \mean{\hat\psi_z}^2 \mean{\delta\hat\psi_z^{\dagger2}} \\
\Rightarrow \dot{\mathcal N} \bar n &= 0.
\end{align}
\end{subequations}
Thus, we conclude that $\partial_t \bar n = \dot{\mathcal D} \bar n + \dot{\mathcal N} \bar n = 0$, so total energy is conserved.

Next, for $\chi^{(2)}$ nonlinear propagation, the total energy is given by the Manley-Rowe invariant
\begin{subequations}
\begin{align}
\bar n_\text{MR} &\coloneqq \bar n_\text{a} + 2\bar n_\text{b},
\quad\text{where} \\
\bar n_\text{a} &\coloneqq \textstyle\int \dif z\, \mean{\hat\phi_z^\dagger\hat\phi_z}
= \int \frac{\dif k}{2\pi}\, \mean{\hat\Phi_k^\dagger\hat\Phi_k} \\
\bar n_\text{b} &= \textstyle\int \dif z\, \mean{\hat\psi_z^\dagger\hat\psi_z} = \int \frac{\dif k}{2\pi}\, \mean{\hat\Psi_k^\dagger\hat\Psi_k},
\end{align}
\end{subequations}
which represents a generalized particle number for the two-envelope model used in Sec.~\ref{sec:chi2-formalism}.
Similarly to the case of $\chi^{(3)}$, one can straightforwardly show that for linear evolution under $\hat H_\text{lin}$ in \eqref{eq:chi2-hamiltonian-continuum},
\begin{equation}
\dot{\mathcal D} \bar n_\text{a} = \dot{\mathcal D} \bar n_\text{b} = \dot{\mathcal D} \bar n_\text{MR} = 0.
\end{equation}
On the other hand, for nonlinear evolution under $\hat H_\text{d3wm}$, we can calculate
\begin{subequations}
\begin{align}
\dot{\mathcal N} \paren{\mean{\hat\phi_z^\dagger}\mean{\hat\phi_z}} &= \mean{\hat\phi_z^\dagger}^2 \mean{\hat\psi_z} + \mean{\hat\phi_z}^2 \mean{\hat\psi_z^\dagger} \\
&\quad {} + \mean{\hat\phi_z^\dagger} \covar{\hat\phi_z^\dagger}{\hat\psi_z} + \mean{\hat\phi_z} \covar{\hat\phi_z}{\psi_z^\dagger}, \nonumber\\
\dot{\mathcal N} \paren{\mean{\hat\psi_z^\dagger}\mean{\hat\psi_z}} &= -\frac{1}{2} \left( \mean{\hat\psi_z^\dagger} \mean{\hat\phi_z}^2 + \mean{\hat\psi_z} \mean{\hat\phi_z^\dagger}^2 \right. \\
&\quad\qquad \left. {}+ \mean{\hat\psi_z^\dagger} \var{\hat\phi_z} + \mean{\hat\psi_z} \mean{\delta\hat\phi_z^{\dagger2}} \right), \nonumber\\
\dot{\mathcal N} \covar{\hat\phi_z^\dagger}{\hat\phi_z} &= \mean{\hat\phi_z} \covar{\hat\phi_z}{\hat\psi_z^\dagger} + \mean{\hat\psi_z^\dagger} \var{\hat\phi_z} \\
&\quad {} + \mean{\hat\phi_z^\dagger} \covar{\hat\phi_z^\dagger}{\hat\psi_z} + \mean{\hat\psi_z} \mean{\delta\hat\phi_z^{\dagger2}}, \nonumber\\
\dot{\mathcal N} \covar{\hat\psi_z^\dagger}{\hat\psi_z} &= -\mean{\hat\phi_z^\dagger} \covar{\hat\phi_z^\dagger}{\hat\psi_z} - \mean{\hat\phi_z} \covar{\hat\phi_z}{\psi_z^\dagger},
\end{align}
\end{subequations}
from which we obtain $ \dot{\mathcal N} \bar n_\text{MR} = 0$ (though $\dot{\mathcal N} \bar n_\text{a}, \dot{\mathcal N} \bar n_\text{b} \neq 0$ in general).
Thus, we conclude that $\partial_t \bar n_\text{MR} = \dot{\mathcal D}\bar n_\text{MR} + \dot{\mathcal N}\bar n_\text{MR} = 0$, so total energy is conserved.

\section{Discretization of the field} \label{sec:discretization}

In the main text, for convenience, we treat the quantum fields propagating on a nonlinear waveguide as being continuous, i.e., $\hat\psi_z$ is an annihilation operator at each $z \in \mathbb R$, with commutation relations $\sbrak{\hat\psi_z, \hat\psi_{z'}^\dagger} = \delta(z-z')$.
For numerical simulations, it is more convenient to use a finite set of discrete modes instead of the continuum, so that, e.g., the mean of the field $\mean{\hat\psi_z}$ can be approximated by a vector and the covariance $\covar{\hat\psi_z}{\hat\psi_{z'}}$ by a matrix.

To define these discrete modes, we introduce a quantization window of length $L$ large enough to contain the pulse of interest.
Note that under an appropriate rotating frame as used, e.g., in \eqref{eq:chi3-hamiltonian-continuum}, this quantization window can be taken to copropagate at the group velocity of the carrier.
We impose periodic boundary conditions on the window, which therefore supports monochromatic waveforms that are periodic with $L$, corresponding to discrete wavevectors $k_0 + m\Delta k$ ($m \in \mathbb Z$), where $\Delta k \coloneqq 2\pi/L$.
We quantize each of these momentum-space modes by introducing mode annihilation operators $\hat A_m$ for each $m$, satisfying $\sbrak[\big]{\hat A_m, \hat A_n^\dagger} = \delta_{mn}$.

To obtain discrete modes in the spatial domain as well, we also impose a bandwidth limit $-M/2 \leq m < M/2$, where, for convenience, we take $M$ to be an even integer; this corresponds to a momentum cutoff $M\Delta k$.
We can now use the discrete Fourier transform to define finite spatial modes
\begin{subequations}
\begin{equation}
\hat a_i \coloneqq \frac{1}{\sqrt M} \sum_{m=-M/2}^{M/2-1} \e{+2\pi\im mi} \hat A_m,
\end{equation}
from which $\sbrak[\big]{\hat a_i, \hat a_j^\dagger} = \delta_{jk}$.
These modes approximately correspond to spatial bins of size $\Delta z \coloneqq L/M$, with $\hat a_i$ annihilating a mode centered on $z = i\Delta z$, if we take the quantization window to be the interval $[-L/2,L/2)$.
Of course, we also have the inverse relation
\begin{equation}
\hat A_m = \frac{1}{\sqrt M} \sum_{i=-M/2}^{M/2-1} \e{-2\pi\im mi} \hat a_i.
\end{equation}
\end{subequations}
Heuristically, the discrete modes we have defined can be thought of as $\hat a_i \sim \hat\psi_{i\Delta z} \sqrt{\Delta z}$ and $\hat A_m \sim \hat\Psi_{m\Delta k} \sqrt{\Delta k/2\pi}$; that is, they are intuitively ``bin'' modes in the spatial and momentum domains, respectively.
Note that a consequence of this interpretation is that the photon numbers in each of these bin modes, i.e., $\mean{\hat a_i^\dagger\hat a_i}$ or $\mean{\hat A_m^\dagger\hat A_m}$, are not intrinsic quantities, as they depend arbitrarily on the chosen values of $L$ and $M$.

{
\renewcommand{\arraystretch}{1.3}
\setlength{\tabcolsep}{6pt}
\begin{table}[t!]
\begin{tabular}{l l l}
\hline\hline
Quantity & Continuous & Discrete \\
\hline
$z$-space index & $z$ & $i \coloneqq z /\Delta z$ \\
$k$-space index & $k$ & $m \coloneqq k /\Delta k$ \\
$z$-space field(s) & $\cbrak{\hat\phi_z, \hat\psi_z}$ & $\cbrak{\hat b_i,\hat a_i} (\Delta z)^{-1/2}$ \\
$k$-space field(s) & $\cbrak{\hat\Phi_k, \hat\Psi_k}$ & $\cbrak{\hat B_m,\hat A_m}\paren[\big]{\frac{\Delta k}{2\pi}}^{-1/2}$ \\
$z$-space measure & $\int\dif z$ & $\sum_i \Delta z$ \\
$k$-space measure & $\int\dif k$ & $\sum_m \Delta k$ \\
\hline\hline
\end{tabular}
\caption{
Formal mappings between quantities in the continuous and discrete descriptions of the multimode nonlinear Gaussian-state model.
The continuous description is used throughout the main text, but for numerical simulations it is useful to convert to the discrete description.
In doing so, we introduce simulation parameters $L$ and $M$, representing the quantization window length and number of sample points (i.e., momentum bandwidth), respectively; these parameters determine $\Delta z \coloneqq L/M$ and $\Delta k \coloneqq 2\pi/L$.
For the summations, both $i$ and $m$ range from $-M/2$ to $M/2-1$, for $M$ even.
} \label{tab:discretization}
\end{table}
}

With these definitions, we can convert continuum field operators and their Gaussian moments to their corresponding discretized quantities.
Table \ref{tab:discretization} gives a formal way to map from the continuous quantities presented in the main text to discrete quantities that are more suitable for numerical simulation.
To illustrate, this procedure produces the following discrete representation of the $\chi^{(3)}$ Hamiltonian $\hat H_{\chi^{(3)}} = \hat H_\text{4wm} + \hat H_\text{lin}$ from \eqref{eq:chi3-hamiltonian-continuum}:
\begin{subequations}
\begin{align}
\hat H_\text{4wm} &= \frac{\hbar}{2} \sum_{i=-M/2}^{M/2-1} \frac{g}{\Delta z} \hat a_i^{\dagger2} \hat a_i^2, \\
\hat H_\text{lin} &= \hbar \sum_{m=-M/2}^{M/2-1} \Omega(m\Delta k) \hat A_m^\dagger \hat A_m,
\end{align}
\end{subequations}
which generates the discretized GSSF equations
\begin{subequations}
\begin{align}
\dif\mathcal N \hat a_i &= -\im\frac{g}{\Delta z} \hat a_i^\dagger \hat a_i^2 \,\dif t, \\
\dif\mathcal D \hat A_m &= -\im\Omega(m\Delta k) \hat A_m \, \dif t. \label{eq:chi3-linear-eom-discrete}
\end{align}
\end{subequations}
Therefore, the GSSF method for a $\chi^{(3)}$ waveguide can be formulated in terms of the finite dynamical quantities $\mean{\hat a_i}$, which is an $M$-dimensional vector, and $\covar{\hat a_i}{\hat a_j}$ and $\covar{\hat a_i^\dagger}{\hat a_j}$, which are $(M\times M)$-dimensional matrices.

\section{Linear losses in waveguide propagation} \label{sec:dissipation}

In the main text, we view the propagation of light through a waveguide as being lossless, in that the dynamics are generated by a Hamiltonian which conserves energy and therefore only incorporates nonlinearity and dispersion.
However, all realistic waveguides feature some amount of propagation loss which occurs continuously throughout the evolution of the pulse in the waveguide.
Since these loss mechanisms usually arise from distributed and disordered effects such as scattering (from surface roughness, etc.), they are well characterized as a source of decoherence of the quantum state.
One approach to modeling such mechanisms is to consider the propagation loss as being analogous to linear dissipation of a generic optical system coupled to a Markovian reservoir, allowing us to use open-quantum-systems theory to treat the effect of this decoherence on our Gaussian state.
Specifically, we posit that the evolution of the density matrix $\hat\rho$ of the system is described not by the Schr\"odinger equation $\partial_t \hat\rho = -(\im/\hbar)\sbrak{\hat H_\text{nl}, \hat\rho}$ (where $\hat H_\text{nl}$ can be $\hat H_{\chi{(3)}}$, $\hat H_{\chi^{(2)}}$, etc.), but rather a master equation in Lindblad form.

We first consider the case of $\chi^{(3)}$ nonlinear propagation and use the discrete description of the field given in Appendix~\ref{sec:discretization}.
Then a suitable quantum master equation for modeling multimode linear losses in the waveguide is
\begin{subequations} \label{eq:loss-master-equation}
\begin{equation}
\partial_t \hat\rho = \frac{1}{\im\hbar} \sbrak{\hat H_\text{nl}, \hat\rho} + \sum_m \paren[\Big]{\hat L_m \hat\rho \hat L_m^\dagger - \frac 1 2 \cbrak[\big]{\hat L_m^\dagger\hat L_m, \hat\rho}},
\end{equation}
where the Lindblad operators $\hat L_m$ represent dissipation in each wavespace mode $m$.
If the mode $\hat A_m$ experiences a field loss rate of $\kappa_m$, then we set
\begin{equation}
\hat L_m \coloneqq \sqrt{2\kappa_m} \hat A_m.
\end{equation}
\end{subequations}
Since the dissipation part is written only in terms of wavespace modes $\hat A_m$, it is clear that loss only affects the Fourier part of the the split-step dynamics in GSSF, generated by $\dif\mathcal D$.

While the master equation \eqref{eq:loss-master-equation} describes evolution of the state in the Schr\"odinger picture, our nonlinear Gaussian-state approximation is best understood in the Heisenberg picture.
As a result, we turn to an equivalent formulation of \eqref{eq:loss-master-equation} in the form of a Heisenberg-Langevin equation of motion.
More formally, the operator $\dif\mathcal D$ formally becomes a stochastic differential propagator, generating a quantum stochastic differential equation~\footnote{In general, under quantum input-output theory, the evolution is governed by a quantum stochastic differential equation (i.e., a Heisenberg-Langevin equation) via $\dif\hat x = \frac{\im}{\hbar} \sbrak{\hat H, \hat x} \,\dif t + \frac{1}{2}\sum_\ell \paren*{\hat L_\ell^\dagger \sbrak{\hat x, \hat L_\ell} + \sbrak{\hat L_\ell^\dagger, \hat x}\hat L_\ell} \,\dif t + \sum_\ell\paren*{\sbrak{\hat L_\ell^\dagger, \hat x} \,\dif\hat W_\ell + \sbrak{\hat x, \hat L_\ell} \,\dif\hat W_\ell^\dagger}$.}
\begin{equation} \label{eq:chi3-qsde}
\dif\mathcal D \hat A_m
= -\im\widetilde\Omega_m \hat A_m \,\dif t
- \sqrt{2\kappa_m} \,\dif\hat W_m,
\end{equation}
where $\widetilde\Omega_m \coloneqq \Omega(m\Delta k) - \im\kappa_m$, and we have introduced input quantum white-noise operators $\dif\hat W_m$ which obey
\begin{subequations}
\begin{align}
\mean{\dif\hat W_m(t)} &= 0, \label{eq:white-noise-mean} \\
\dif\hat W_m(t) \dif\hat W_{m'}^\dagger(t') &= \delta_{mm'}\delta(t-t') \,\dif t, \label{eq:white-noise-variance}
\end{align}
\end{subequations}
with all other possible products being zero.
In the presence of linear loss, \eqref{eq:chi3-qsde} represents a generalization of \eqref{eq:chi3-linear-eom} and \eqref{eq:chi3-linear-eom-discrete}.

We therefore need to use \eqref{eq:chi3-qsde} to rederive the appropriate equations of motion for $\mean{\hat A_m}$, $\covar{\hat A_m}{\hat A_n}$ and $\covar{\hat A_m^\dagger}{\hat A_n}$ in the Fourier step.
As expected, the mean equation is unaffected by the quantum white-noise term by virtue of \eqref{eq:white-noise-mean}, and it only picks up a field loss:
\begin{equation}
\dif\mathcal D \mean{\hat A_m} = -\im\widetilde\Omega_m \mean{\hat A_m} \,\dif t,
\end{equation}
which generalizes \eqref{eq:chi3-gaussian-mean-linear} for linear loss.
In calculating the equations of motion for the covariances, we note that in general, there is a contribution from $\mean{(\dif\mathcal D\delta\hat A_m)(\dif\mathcal D\delta\hat A_n)}$, as a result of a quantum generalization of It\^o's lemma for the stochastic differentials $\dif\hat W_m$, which would then require simplification using \eqref{eq:white-noise-variance}.
However, because we are focusing on normal-ordered covariances involving the mode operators, such It\^o terms happen to be zero; such terms cannot be neglected, e.g., for quadrature covariances like $\covar{\hat q_m}{\hat q_n}$, where $\hat q_m \coloneqq \frac{1}{\sqrt2}(\hat A_m + \hat A_m^\dagger)$.
In the end, we find that
\begin{subequations}
\begin{align}
\dif\mathcal D\covar{\hat A_m}{\hat A_n} &= -\im\paren[\big]{\widetilde\Omega_n + \widetilde\Omega_m} \covar{\hat A_m}{\hat A_n} \,\dif t, \\
\dif\mathcal D\covar{\hat A_m^\dagger}{\hat A_n} &= -\im\paren[\big]{\widetilde\Omega_n - \widetilde\Omega_m^*} \covar{\hat A_m^\dagger}{\hat A_n} \,\dif t,
\end{align}
\end{subequations}
which generalizes \eqref{eq:chi3-gaussian-covars-linear} for linear loss.
The same calculations can be done for $\chi^{(2)}$ waveguide propagation.

\section{Supermode decomposition} \label{sec:supermode}

In many of our examples, we would like to find squeezing supermodes using the Gaussian moments of the fields that we simulate.
Here, we describe the construction of the standard covariance matrix $\Sigma$ written in terms of quadrature, rather than mode, operators, as well as how to use $\Sigma$ to calculate the squeezing supermodes of the multimode fields.

Consider a set of mode operators $\hat c_i$ ($1 \leq i \leq N$) such that $\sbrak{\hat c_i, \hat c_j^\dagger} = \delta_{ij}$.
Here, $\hat c$ could consist of, e.g., discretized modes of a continuous field as described in Appendix~\ref{sec:discretization}, and we note that $\hat c_i$ can also include combinations of fields as well, e.g., $\hat c = (\hat a_1, \ldots, \hat a_M, \hat b_1, \ldots, \hat b_M)$.
We first define $\hat q_i \coloneqq \frac{1}{\sqrt2} \paren[\big]{\hat c_i + \hat c_i^\dagger}$ and $\hat p_i \coloneqq \frac{1}{\sqrt2\im}\paren[\big]{\hat c_i - \hat c_i^\dagger}$ to be the in-phase (real) and quadrature-phase (imaginary) components of $\hat c_i$, respectively.
Then, the standard covariance matrix in quadrature form is defined as $\Sigma_{k\ell} \coloneqq \mean{\textstyle\frac12\cbrak{\delta\hat z_k, \delta\hat z_\ell}}$ where $\hat z \coloneqq (\hat q_1, \ldots, \hat q_N, \hat p_1, \ldots, \hat p_N)$, and $\cbrak{\hat z_1, \hat z_2} \coloneqq \hat z_1 \hat z_2 + \hat z_2 \hat z_1$ is the anticommutator.

In general, $\Sigma$ is a $2N \times 2N$ positive-definite matrix on which we can perform a Williamson decomposition $\Sigma = SDS^\mathrm{T}$ where $D$ is a diagonal matrix of the form $D = \frac 1 2 + \diag(\bar n, \bar n)$ where $\bar n_i \geq 0$ and $S$ is a symplectic matrix satisfying $S\Omega S^\mathrm{T} = \Omega$, for $\Omega \coloneqq \begin{psmallmatrix} 0 & 1_N \\ -1_N & 0 \end{psmallmatrix}$ the symplectic form.
Physically, $S$ represents a set of quantum-limited multimode phase-shifting, mode-mixing, and squeezing operations acting on an $N$-mode initial thermal state with thermal photon populations $\bar n_i$.

We can furthermore perform a Bloch-Messiah (or Euler) decomposition $S = O_\text{out}\Lambda O_\text{in}$ where $O_\text{out}$ and $O_\text{in}$ are orthogonal symplectic matrices and $\Lambda$ is a diagonal matrix of the form $\Lambda = \diag(\e{-r_1},\ldots,\e{-r_N},\e{+r_1},\ldots,\e{+r_N})$.
(Note that this convention does not list the elements in increasing order.)
Physically, the $N$-mode operation represented by $S$ is being decomposed into a set of (active) single-mode squeezers with squeezing parameters $r_i$, sandwiched between an input set of $N$-mode (passive) beamsplitters and phaseshifters represented by $O_\text{in}$ and a similar set at the output represented by $O_\text{out}$.

Since $O_\text{out}$ passively generates multimode squeezing from single-mode squeezing, it is also the matrix that determines the separable supermodes of the Gaussian state.
In general, $O_\text{out}$ (and $O_\text{in}$) have the general form $\begin{psmallmatrix} X & -Y \\ Y & X \end{psmallmatrix}$, and $U \coloneqq X + \im Y$ is a $N \times N$ unitary matrix, which defines the supermodes via
\begin{equation}
\hat C_i \coloneqq \sum_{j=1}^N U_{ij} \hat c_j.
\end{equation}
Note that while they are separable, these supermodes are not necessarily uncorrelated; they have a covariance matrix $\widetilde\Sigma \coloneqq O_\text{out}^\mathrm{T} \Sigma O_\text{out} = \Lambda O_\text{in} D O_\text{in}^{\mathrm{T}} \Lambda$, which is diagonal if $D \propto 1_{2N}$; the off-diagonal elements represent correlations due to multi-(super)mode mixtures of thermal photons.
For a pure Gaussian state, $D = \diag(1/2)$, so the first $N$ diagonal elements of $\widetilde\Sigma$ give the variance along the squeezed quadrature of each supermode $\hat C_i$, respectively, while the last $N$ diagonal elements give the corresponding variance along the antisqueezed quadratures; for $\hat C_i$, these variances are $\frac12\e{\pm 2r_i}$, respectively.

\section{Gaussian theory for self-heterodyne detection of CEO beat note} \label{sec:heterodyne}

In this appendix, we derive expressions for the signal and noise of an $f-2f$ beat note obtained by self-heterodyning a supercontinuum frequency comb consisting of a fundamental-harmonic (FH) and a second-harmonic (SH) envelope, when the state of the field is in a multimode-entangled Gaussian state.
In self-heterodyning, we have overlapping frequency components between the FH and SH envelopes which interfere at a photodetector to produce a heterodyne signal.
This situation is slightly different from the usual quantum-optical setup for single-mode heterodyne involving a strong local oscillator in a coherent state.
Nevertheless, expressions for the signal and noise of the result can be derived using the same photodetection theory.
Here, we follow the formalism of Ref.~\cite{Collett1987} to do so.

The heterodyne photocharge received after demodulating the photocurrent with an electronic local oscillator at frequency $f_\text{h}$ and then integrating for time $T_\text{h}$ (assumed larger than the pulse duration), can be expressed as~\cite{Collett1987}
\begin{equation} \label{eq:Qh-mean}
\mean*{Q_\text{h}} = \int_0^{T_\text{h}} \! v_\text{h}(t) G_1(t) \,\dif t,
\end{equation}
where $v_\text{h}(t) \coloneqq \cos(2\pi f_\text{h}t)$ and the first-order correlation function is
\begin{equation}
G_1(t) \coloneqq \mean[\big]{\hat D^\dagger(t) \hat D(t)},
\end{equation}
for some appropriate choice of $\hat D$ such that $\hat D^\dagger \hat D$ captures the photon flux at the surface of the detector.

On the other hand, the noise on that photocharge has a variance given by~\cite{Collett1987}
\begin{align} \label{eq:Qh-variance}
\mean[\big]{\delta Q_\text{h}^2} &= \int_0^{T_\text{h}} \! v_\text{h}^2(t) G_1(t) \,\dif t \\
&\qquad{}+ \int_0^{T_\text{h}} \! \int_0^{T_\text{h}} \! v_\text{h}(t) v_\text{h}(t') G_2(t,t') \,\dif t\,\dif t',\nonumber
\end{align}
where the first term represents the shot noise associated to $\mean*{Q_\text{h}}$ and the second term is contributed by the second-order correlation function
\begin{align}
G_2(t,t') &\coloneqq \mean[\big]{\hat D^\dagger(t) \hat D^\dagger(t') \hat D(t) \hat D(t')} - G_1(t)G_1(t').
\end{align}

To develop $G_1$ and $G_2$ further, we specialize to a pulsed field generated by a frequency comb with repetition rate $f_\text{rep}$.
We assume the field consists of a FH envelope with carrier-envelope offset (CEO) frequency $f_\text{ceo} < f_\text{rep}$ and an SH envelope with CEO frequency $2f_\text{ceo}$.
The modes of the FH comb, denoted by $\hat A_m$, have frequency $(m+m_0)f_\text{rep} + f_\text{ceo}$, and the modes of the SH comb, denoted by $\hat B_q$, have frequency $(q+2m_0)f_\text{rep} + 2f_\text{ceo}$, where $m_0$ is a fixed integer denoting the central FH comb line, and we have as usual $\sbrak{\hat A_m, \hat A_{m'}^\dagger} = \delta_{mm'}$ and $\sbrak{\hat B_q, \hat B_{q'}^\dagger} = \delta_{qq'}$.
We also assume that the linewidth of the comb lines are much smaller than both $f_\text{ceo}$ and $f_\text{rep} - f_\text{ceo}$, so that two spectral modes with optical frequencies separated only by $f_\text{ceo}$ are distinguishable, so $\sbrak{\hat A_m, \hat B_q^\dagger} = 0$ for all $m,q$.

Let us move to the rotating frame of mode $\hat A_0$ (with the CEO frequency included), so that all mode operators now rotate at $m_0 f_\text{rep} + f_\text{ceo}$.
In this frame, we can construct a field operator (at the surface of the detector) out of the modes $\hat A_m$ and $\hat B_q$ via
\begin{align} \label{eq:field-pulse}
\hat d(t) &= \sum_m \hat A_m \e{-2\pi\im mf_\text{rep} t} \\
&\qquad{}+ \sum_q \hat B_q \e{-2\pi\im(q+m_0)f_\text{rep}t} \e{-2\pi\im f_\text{ceo} t}.\nonumber
\end{align}
However, $\hat d$ does \emph{not} physically describe a frequency comb and its corresponding pulse train.
Rather, the physical scenario described by \eqref{eq:field-pulse} is a \emph{single pulse} obeying a periodic boundary; alternatively, we may interpret \eqref{eq:field-pulse} as a Fourier series decomposition of the \emph{first pulse} of the train.
For example, while the field $\hat d$ has finite energy, a true pulse train would not, even though the flux (evaluated at some reference plane) may be the same.
As we will see, the distinction between these two scenarios is important for calculating multi-time correlation functions such as $G_2(t,t')$.
To fix the issue, we note that since the physical pulse train we are considering does not exhibit any interpulse correlations (i.e., each pulse is the same as the next), and the only difference we have to track is the shift in the CEO phase from one pulse to the next.
In this case, we can take
\begin{equation}
\hat D(t) = \sqrt{f_\text{rep}} \sum_{\ell=-\infty}^{\infty} \Pi(f_\text{rep} t - \ell) \, \hat d_\ell(t),
\end{equation}
where the rectangle function $\Pi(x)$ is one if $0 < x < 1$ and zero otherwise.
By writing $\hat d_\ell$, we formally mean that we substitute in \eqref{eq:field-pulse} $\hat A_m \mapsto \hat A^{(\ell)}_m$ and $\hat B_q \mapsto \hat B^{(\ell)}_q$, which, for different values of the superscript $\ell$, are independent modes but distributed identically to our original wavespace modes $\hat A_m$ and $\hat B_q$.
Formally, let $\hat C^{(\ell)}$ be any product of operators from among $\hat A_m^{(\ell)}, \hat B_q^{(\ell)}$ (and their adjoints), so that any expectation value can be written as $\mean[\big]{\hat C_1^{(\ell_1)}C_2^{(\ell_2)}\cdots}$, where $\ell_1 < \ell_2 < \cdots$.
Then under our ``independent but identically distributed'' condition, $\mean[\big]{\hat C_1^{(\ell_1)} \hat C_2^{(\ell_2)}\cdots} = \mean[\big]{\hat C_1}\mean[\big]{\hat C_2}\cdots$ (without superscripts).

We also note that we have turned $\hat D^\dagger \hat D$ into a flux quantity by normalizing the photons per pulse by the repetition time, which corresponds to assuming a separation between a ``fast timescale'' on the order of the pulse duration (e.g., $\sim \SI{100}{fs}$), and a ``slow timescale'' on the order of the repetition time (e.g., $\sim\SI{1}{ns}$).

Using this form for $\hat D(t)$, we can now calculate that
\begin{align}
G_1(t) &= f_\text{rep} \sum_{l=-\infty}^{\infty} \Pi(f_\text{rep}t - \ell) \, \mean[\big]{\hat d_\ell^\dagger(t) \hat d_\ell(t)} \nonumber\\
&= f_\text{rep} \sum_{\ell=-\infty}^\infty \Pi(f_\text{rep}t - \ell) \, g_1(t). \label{eq:G1-comb}
\end{align}
where, in the first line, we have used $\Pi(x-\ell)\Pi(x-r) = \Pi(x-\ell)\delta_{lr}$, and in the second line used the fact that the expectation value only involves operators from the same pulse index $\ell$, which allows us to express the result in terms of the first-order correlation function for a ``single pulse'', defined as
\begin{equation}
g_1(t) \coloneqq \mean[\big]{\hat d^\dagger(t) \hat d(t)}.
\end{equation}
Using similar arguments, we find that
\begin{align}
&G_2(t,t') = f_\text{rep}^2 \sum_{\ell,\ell'=-\infty}^\infty \Pi(f_\text{rep}t - \ell) \Pi(f_\text{rep}t' - \ell') \nonumber\\
&\qquad{}\times \mean[\big]{\hat d_\ell^\dagger(t) \hat d_{\ell'}^\dagger(t') \hat d_\ell(t) \hat d_{\ell'}(t')} - G_1(t) G_1(t') \nonumber \\
&= f_\text{rep}^2 \sum_{\ell=-\infty}^\infty \Pi(f_\text{rep} t - \ell) \Pi(f_\text{rep} t' - \ell) \, g_2(t,t'), \label{eq:G2-comb}
\end{align}
where for the second line we used the fact that the subtraction of $G_1(t)G_1(t')$ eliminates all terms of the sum for which $\ell \neq \ell'$, allowing us to similarly express the result in terms of the second-order correlation function for a single pulse, defined as
\begin{equation}
g_2(t,t') \coloneqq \mean[\big]{\hat d^\dagger(t) \hat d^\dagger(t') \hat d(t) \hat d(t')} - g_1(t) g_1(t').
\end{equation}
It is worth emphasizing the form of \eqref{eq:G2-comb}, in which having two windows tied to the same index $\ell$ ensures that $g_2(t,t')$ \emph{does not contribute} to $G_2(t,t')$ when $|t-t'| > 1/f_\text{rep}$.
This is essential for enforcing the independence of the modes constituting each individual pulse, despite the fact that we are able to write the result in terms of only the single-pulse modes $\hat A_m$ and $\hat B_q$ due to the quasi-periodicity of the pulse train.

At this point, we have finished setting up the model, and we can proceed with the calculation of the heterodyne signal $\mean*{Q_\text{h}}$ and noise $\mean[\big]{\delta Q_\text{h}^2}$.
In doing so, we set $f_\text{h} = f_\text{ceo}$ and take $T_\text{h} \gg 1/f_\text{ceo} > 1/f_\text{rep}$ (but still smaller than the coherence time of each comb line), so that all beat notes with frequency larger than $f_\text{ceo}$ wash out and we are left only with the components at the beat note $f_\text{ceo}$, demodulated to DC.
We also take $T_\text{h}$ to be an integer multiple of the pulse repetition time $1/f_\text{rep}$, which causes no loss of generality in the limit $T_\text{h} \gg 1/f_\text{rep}$.

We first calculate $\mean*{Q_\text{h}}$.
Inserting \eqref{eq:G1-comb} into \eqref{eq:Qh-mean},
\begin{equation} \label{eq:Qh-temp}
\mean*{Q_\text{h}} = \sum_{\ell=0}^{T_\text{h}f_\text{rep}} \int_{\ell}^{\ell+1}\!\! \dif s\, \cos(\phi_\text{ceo}s) \, g_1(t),
\end{equation}
where we have changed integration variables to $s = tf_\text{rep}$ and defined the CEO phase $\phi_\text{ceo} \coloneqq 2\pi f_\text{ceo}/f_\text{rep}$.
The $\Pi(f_\text{rep}t-\ell)$ windows have broken up the integration over $t$ into a sum of integrals over individual pulses.
In the limit $T_\text{h} \gg 1/f_\text{rep}$, we can evaluate the integrals in \eqref{eq:Qh-temp} at each $\ell$, and then neglect all oscillating terms in $\ell$ that do not scale with $T_\text{h} f_\text{rep}$.
Furthermore, we can restrict our attention to only those terms in $g_1(t)$ with oscillations at $f_\text{ceo}$, which are
\begin{align*}
&\sum_{m,q} \mean[\big]{\hat A_m^\dagger \hat B_q} \e{-2\pi\im(q-m+m_0)f_\text{rep}t} \e{-2\pi\im f_\text{ceo}t} + \text{c.c.}
\end{align*}
However, we clearly also require $m_0 + q - m = 0$, which physically means that only interference from FH and SH lines that are next to each other (separated by $f_\text{ceo}$) contribute.
For convenience, let us introduce the shorthand $q(m) \coloneqq m - m_0$.
Inserting only the relevant terms of $g_1(t)$ into \eqref{eq:Qh-temp}, evaluating the sum of integrals, and applying the limit $T_\text{h}f_\text{rep} \gg 1$ we get
\begin{equation}
\mean*{Q_\text{h}} = T_\text{h}f_\text{rep} \sum_m \RE\mean[\big]{\hat A_m^\dagger \hat B_{q(m)}}.
\end{equation}
If we now define the steady-state photocurrent to be $\mean*{I_\text{h}} \coloneqq \mean*{Q_\text{h}}/T_\text{h}$, then we exactly get the expression \eqref{eq:fceo-signal} from the main text.

The calculation for $\mean[\big]{\delta Q_\text{h}^2}$ is similar.
Inserting \eqref{eq:G2-comb} and \eqref{eq:G1-comb} into \eqref{eq:Qh-variance},
\begin{align} \label{eq:Qh2-temp}
&\mean[\big]{\delta Q_\text{h}^2} = \sum_{\ell=0}^{T_\text{h}f_\text{rep}} \int_{\ell}^{\ell+1}\!\! \cos^2(\phi_\text{ceo}s) \, g_1(t) \,\dif s \\
&\;{}+ \sum_{\ell=0}^{T_\text{h}f_\text{rep}} \int_\ell^{\ell+1}\!\!\!\int_\ell^{\ell+1}\!\! \cos(\phi_\text{ceo}s)\cos(\phi_\text{ceo}s') \, g_2(t,t') \,\dif s\,\dif s', \nonumber
\end{align}
where, again, $t = s/f_\text{rep}$ and $t' = s'/f_\text{rep}$.
This time, the first integral only picks out components of $g_1(t)$ oscillating at DC.
These terms are
\begin{equation*}
\sum_m \paren*{\mean[\big]{\hat A_m^\dagger \hat A_m} + \mean[\big]{\hat B_m^\dagger \hat B_m}}.
\end{equation*}
The second integral is more intensive: the relevant terms of $g_2(t,t')$ are those whose time dependence takes the form $\e{-2\pi\im f_\text{ceo}(t \pm t')}$, which consists of the terms
\begin{widetext}
\begin{align*}
&\sum_{m,m'}
\sbrak*{\paren*{
\mean[\big]{\hat A_m^\dagger \hat A_{m'}^\dagger \hat B_{q(m)} \hat B_{q(m')}}
- \mean[\big]{\hat A_m^\dagger \hat B_{q(m)}} \mean[\big]{\hat A_{m'}^\dagger \hat B_{q(m')}}
} \e{-2\pi\im f_\text{ceo}(t+t')}
+ \text{c.c.}
} \\
&\quad{}+ \sum_{m,m'}
\sbrak*{\paren*{
\mean[\big]{\hat A_m^\dagger \hat B_{q(m')}^\dagger \hat A_{m'} \hat B_{q(m)}}
- \mean[\big]{\hat A_m^\dagger \hat B_{q(m)}} \mean[\big]{\hat A_{m'} \hat B_{q(m')}^\dagger}
} \e{-2\pi\im f_\text{ceo}(t-t')}
+ \text{c.c.}
}.
\end{align*}
\end{widetext}
We then need to insert all these terms for both $g_1(t)$ and $g_2(t,t')$ into \eqref{eq:Qh2-temp}, evaluate the sums of integrals, apply the limit $T_\text{h}f_\text{rep} \gg 1$, and perform some algebra.
The end result is that we can define the variance of the photocurrent to be $\mean*{\delta I_\text{h}^2} \coloneqq \mean[\big]{\delta Q_\text{h}^2}/T_\text{h}^2$, which is given by \eqref{eq:fceo-variance} in the main text.
It is interesting to note it turns out that the second term depends continuously on $\phi_\text{ceo}$, while the first term does not (except for some discrete, edge cases such as $\phi_\text{ceo} = \pi/2, \pi$, etc., which we neglect).

We end this section with a remark on evaluating the fourth-order moments that show up in \eqref{eq:fceo-variance}.
For a Gaussian state, they can always be simplified into sums over products of second-order moments, using the relationship
\begin{align}
&\mean*{\hat x \hat u \hat v \hat y} - \mean*{\hat x \hat y} \mean*{\hat u \hat v} \\
&\quad{}= \paren[\big]{\mean*{\hat x} \mean*{\hat u} + \covar*{\hat x}{\hat u}} \covar*{\hat v}{\hat y}
+ \mean*{\hat v} \mean*{\hat y} \covar*{\hat x}{\hat u} \nonumber \\
&\qquad\quad{}+ \paren[\big]{\mean*{\hat x} \mean*{\hat v} + \covar*{\hat x}{\hat v}} \covar*{\hat u}{\hat y}
+ \mean*{\hat u} \mean*{\hat y} \covar*{\hat x}{\hat v}. \nonumber
\end{align}

\end{appendix}

\end{document}